\documentclass[lettersize,journal]{IEEEtran}
\usepackage{amsmath,amsfonts}
\usepackage{algorithmic}
\usepackage{algorithm}
\usepackage{array}
\usepackage[caption=false,font=scriptsize,labelfont=sf,textfont=sf]{subfig}
\usepackage{textcomp}
\usepackage{stfloats}
\usepackage{url}
\usepackage{verbatim}
\usepackage{graphicx}
\usepackage{cite}
\usepackage{booktabs}
\usepackage{makecell}
\usepackage{multirow}
\usepackage{xcolor}
\usepackage{orcidlink}

\begin{document}

\title{Generative Latent Coding for Ultra-Low Bitrate Image and Video Compression}

\author{Linfeng Qi\orcidlink{0009-0006-4278-9400},
Zhaoyang Jia\orcidlink{0000-0001-8814-9691},
Jiahao Li\orcidlink{0000-0002-9881-2124}, 
Bin Li\orcidlink{0000-0002-5635-2916}, ~\IEEEmembership{Member,~IEEE}, 
Houqiang Li\orcidlink{0000-0003-2188-3028}, ~\IEEEmembership{Fellow,~IEEE}, 
Yan Lu\orcidlink{0000-0001-5383-6424},
~\IEEEmembership{Member,~IEEE}

        
\thanks{Linfeng Qi, Zhaoyang Jia and Houqiang Li are with MoE Key Laboratory of Brain-inspired Intelligent Perception and Cognition, University of Science and Technology of China (e-mail: qlf324@mail.ustc.edu.cn; jzy\_ustc@mail.ustc.edu.cn; lihq@ustc.edu.cn).}
\thanks{Jiahao Li, Bin Li, and Yan Lu are with Microsoft Research Asia, Beijing 100080, China (e-mail: li.jiahao@microsoft.com; libin@microsoft.com;
yanlu@microsoft.com).
}
\thanks{This work was done when Linfeng Qi and Zhaoyang Jia were interns at Microsoft Research Asia.}
}

\markboth{Qi \MakeLowercase{et al.}: Generative Latent Coding for Ultra-Low Bitrate Image and Video Compression}%
{Qi et al.: Generative Latent Coding for Ultra-Low Bitrate Image and Video Compression}




\maketitle

\IEEEpubid{
\begin{minipage}{\textwidth}
\centering
Copyright~\copyright~2025 IEEE. Personal use of this material is permitted.  However, permission to use this material for any other purposes must be obtained from \\ the IEEE by sending an email to pubs-permissions@ieee.org.
\end{minipage}
}
\IEEEpubidadjcol

\begin{abstract}
Most existing approaches for image and video compression perform transform coding in the pixel space to reduce redundancy. 
However, due to the misalignment between the pixel-space distortion and human perception, such schemes often face the difficulties in achieving both high-realism and high-fidelity at ultra-low bitrate.
To solve this problem, we propose \textbf{G}enerative \textbf{L}atent \textbf{C}oding (\textbf{GLC}) models for image and video compression, termed GLC-image and GLC-Video. The transform coding of GLC is conducted in the latent space of a generative vector-quantized variational auto-encoder (VQ-VAE). Compared to the pixel-space, such a latent space offers greater sparsity, richer semantics and better alignment with human perception, and show its advantages in achieving high-realism and high-fidelity compression.
To further enhance performance, we improve the hyper prior by introducing a spatial categorical hyper module in GLC-image and a spatio-temporal categorical hyper module in GLC-video.
Additionally, the code-prediction-based loss function is proposed to enhance the semantic consistency.
Experiments demonstrate that our scheme shows high visual quality at ultra-low bitrate for both image and video compression.
For image compression, GLC-image achieves an impressive bitrate of less than $0.04$ bpp, achieving the same FID as previous SOTA model MS-ILLM while using $45\%$ fewer bitrate on the CLIC 2020 test set. For video compression, GLC-video achieves 65.3\%  bitrate saving over PLVC in terms of DISTS.

\end{abstract}


\section{Introduction}
\IEEEPARstart{I}{mage} and video compression are crucial in addressing the massive amount of digital visual data that must be stored and transmitted. Most existing approaches, including the traditional codecs~\cite{wallace1991jpeg, bross2021overview} and the emerging learned based codecs~\cite{balle2018variational, cheng2020learned, 9455349, guo2022evc, liu2023learned, fvc, DCVC, 10596319, 10003249}, perform transform coding in the pixel-space ~\cite{balle2017end} to reduce redundancy. Specifically, these paradigms transforms the pixels of the input image or video frame into compact representations,  which are then encoded into bitstreams with fewer bits.

\begin{figure}[t]
  \centering
    \includegraphics[width=\linewidth]{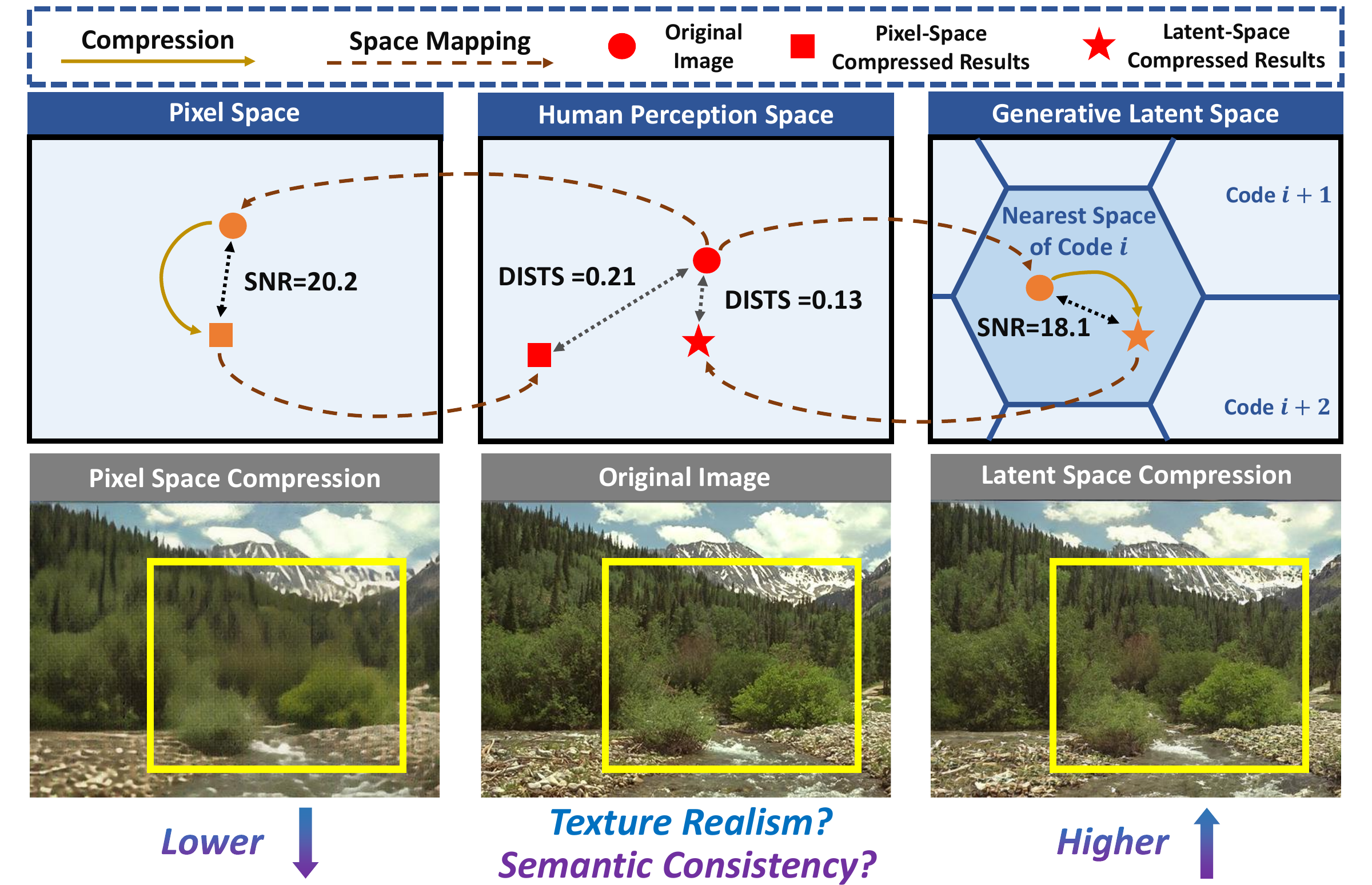}
    \caption{For ultra-low bitrates, the generative latent space of VQ-VAE provides a better alignment with human perception than the pixel space.
    At comparable distortion levels, latent-space compression yields reconstructions with superior perceptual quality compared to the pixel-space generative codec MS-ILLM~\cite{muckley2023improving}, as measured by signal-to-noise ratio (SNR). The perceptual enhancement is quantified using the DISTS metric~\cite{ding2020image}.}
  \label{fig:Pixel_vs_Latent}
  \vspace{-3mm}
\end{figure}

However, we observe a common inherent limitation in these methods: the pixel-space distortion is not always consistent with human perception, especially at low bitrates. Human perception prioritizes semantic consistency and texture realism, which are difficult to be adequately captured using only a pixel-space transform module.
As shown in the left of Fig.~\ref{fig:Pixel_vs_Latent}, MS-ILLM~\cite{muckley2023improving}, previous SOTA pixel-space generative image codec, struggles to guarantee visual quality at low bitrate, even though it is equipped with the perceptual supervision~\cite{johnson2016perceptual} and adversarial supervision~\cite{goodfellow2014generative} within the pixel space.

\IEEEpubidadjcol

To address this limitation, we propose \textbf{G}enerative \textbf{L}atent \textbf{C}oding paradigm for image compression (\textbf{GLC}-image) and video compression (\textbf{GLC}-video).
In our method, input images or video frames are firstly encoded into a generative latent space aligned with human perception, followed by transform coding on the encoded latents to facilitate compression.
Specifically, we utilize a generative vector-quantized variational auto-encoder (VQ-VAE)~\cite{van2017neural, esser2021taming} to learn a generative latent space, which offers three significant advantages: 1) The discrete codes of VQ-VAE encapsulate semantic visual components~\cite{van2017neural}, allowing our model to prioritize compressing semantic content, thereby achieving improved visual quality at ultra-low bitrates. 2) Generative VQ-VAE exhibits remarkable generative capabilities~\cite{esser2021taming} for generating high-realism textures. 3) A low-entropy and distortion-robust latent space can be achieved through the discrete variational bottleneck, which is well-suited for compression tasks. These characteristics align our model more closely with human perception and improve visual quality, as illustrated in the right of Fig.~\ref{fig:Pixel_vs_Latent}.

During implementation, two critical problems need to be addressed: \textit{1. How to effectively compress the generative latents? 2. How to supervise the generative latent coding pipeline?} A straightforward approach to compress the latents of VQ-VAE is indices-map coding~\cite{jiang2023adaptive, mao2023extreme}. 
However, its ineffective redundancy reduction between indices and the lack of supporting rate-variable coding hinder its practical applicability.
To address these limitations, we propose a novel generative-latent-space transform coding approach for GLC-image and GLC-video. The input images or video frames are transformed into latents, and an effective rate-variable structure is implemented to reduce latent redundancy for achieving higher compression ratio. We observe that temporal correlations in video are not restricted to frame pixels but are also preserved within the generative latent space. Leveraging this insight, conditional coding is utilized to compress the current latent using the previously decoded latent as references, effectively reducing temporal redundancy.

In considering effective supervision to guide the training, we draw inspiration from recent advancements in code prediction transformers~\cite{zhou2022towards, jiang2023adaptive}. We propose a code-prediction-based supervision mechanism, which acts as auxiliary supervision employed solely in the training process and greatly enhances the semantic consistency. To further boost the performance, we improve the hyper module for GLC-image and GLC-video. 
For GLC-image, we design the spatial categorical hyper module to address the limitations of the factorized hyper module~\cite{balle2018variational}. At ultra-low bitrates, factorized hyper module primarily allocates bits to low-level features. In contrast, our spatial categorical hyper module employs a discrete codebook to capture basic semantic elements for each spatial positions. It brings a better trade-off between preserving essential semantic details and reducing the bit cost of coding hyper-information.
For GLC-video, we emphasize the necessity of integrating temporal correlations into the hyper prior to reduce redundancy. We also adopt the concept of the spatial categorical hyper module in GLC-image, utilizing a discrete codebook to model the hyper information. However, independently encoding such categorical hyper information across frames overlooks the inherent temporal correlations, resulting in significant redundancy. Temporal context provides a valuable opportunity to predict regions that are more likely to change within a frame, allowing us to focus on encoding the spatio-temporal information. 
Moreover, in video sequences, the dynamics of correlated content often follow regular patterns, but the hyper-features with a local receptive field lacks the capability for capturing global and high-level dynamics, such as background motion or camera perspective changes.
Therefore, we propose the spatio-temporal categorical hyper module. This module aggregates global semantic dynamics into a compact set of tokens, where each token encapsulates the context of the entire frame rather than a localized region. These tokens act as hyper information, significantly reducing redundancy through capturing global patterns and semantic correlations.

By incorporating these advanced designs, our GLC-image and GLC-video models deliver outstanding performance in image and video compression tasks. On the CLIC 2020 test set~\cite{toderici2020clic}, GLC-image achieves a remarkable bitrate of less than $0.04$ bpp while maintaining high visual quality, demonstrating a $45\%$ bit savings compared to MS-ILLM~\cite{muckley2023improving} at an equivalent FID. 
In video compression, GLC-video achieves significantly lower bitrates (down to $0.01$ bpp) on benchmark video datasets compared to previous generative video codecs such as PLVC~\cite{yang2022perceptual}. GLC-video can produce visually superior reconstructions compared to the advanced neural video codec DCVC-FM~\cite{DCVC-FM} and PLVC even with a lower bpp, further highlighting the effectiveness of our proposed approach.

In summary, our main contributions are:

\begin{itemize}
    \item We propose a novel generative latent coding scheme, performing transform coding to reduce latent redundancy within the generative latent space of a VQ-VAE. Our scheme not only supports rate-variable image and video compression, but also results in high-fidelity and realistic reconstructions. 

    \item For GLC-image, we introduce a spatial categorical hyper module to reduce the bit cost of hyper information. For GLC-video, we propose a spatio-temporal categorical hyper module to capture global semantic dynamics to reduce redundancy.

    \item We propose a code-prediction-based supervision strategy to guide the training and tap the potential of the generative latent coding pipeline.
    
    \item GLC-image obtains a 45\% bit reduction on CLIC2020 with the same FID as the previous advanced method MS-ILLM, and GLC-video provides a 65.3\% bitrate saving over PLVC in terms of DISTS on benchmark video datasets, demonstrating the effectiveness of our approach. 
\end{itemize}

This work is built upon our preliminary conference paper~\cite{jia2024generative}, with notable enhancements summarized as follows: 1. Leveraging the success of GLC-image and the insight that temporal correlations remain significant in the latent space, we extend the generative latent coding approach to video compression, introducing the GLC-video framework for achieving ultra-low bitrate video compression. 2. To address temporal redundancy, we design a conditional transform coding module that compresses the current latent by utilizing the previously decoded latent as temporal context, thereby improving coding efficiency.
3. Recognizing the temporal redundancy and global motion patterns inherent in hyper information across video frames, we propose a spatio-temporal categorical hyper module. This module predicts regions of change using temporal context and captures global semantic dynamics, enabling more efficient encoding and improved compression performance. 4. We incorporate more recent methods, such as DiffEIC~\cite{DiffEIC}, for comprehensive comparisons with GLC-image on benchmark datasets. Extensive experiments and analyses are conducted to evaluate the proposed GLC-video framework, further highlighting the advantages of our generative latent coding.

\section{Related work}

\subsection{Learned Image Compression}
Learned image compression has seen rapid advancements in recent years.
Ball\'{e} et al.~\cite{balle2017end} proposed utilizing neural networks for pixel-space transform coding. They introduced transform modules to convert images into compact representations for entropy coding. 
Following this, significant advancements have been made: some studies focus on improving the probability model~\cite{balle2017end, lee2018context,cheng2020learned, balle2018variational, li2023neural} to achieve more accurate estimation, while others enhance network structures~\cite{cheng2020learned, liu2023learned}, refine optimization algorithms~\cite{zhao2023universal} or explore rate-variable coding~\cite{guo2022evc, cui2021asymmetric}. These works improve both compression performance and practical applicability.

A critical challenge in modern image compression lies in improving the perceptual quality of reconstructed images. 
Agustsson et al.~\cite{agustsson2019generative} introduced the concept of \textit{generative compression}, which focuses on compressing essential image features and generating distorted details using generative adversarial networks (GANs). Building on this concept, subsequent works~\cite{hu2020towards, lei2023text+} explored extracting image sketches and latent codes to maintain geometric consistency. However, while these approaches produce visually appealing results, they often deviate significantly from the input, failing to ensure semantic consistency in the reconstructions.

To achieve high-fidelity generative compression, Mentzer et al.~\cite{mentzer2020high} explored advanced network structures and generative adversarial loss to improve the fidelity of the reconstructions. Subsequent methods have made notable advancements, including enhancing the transform coding~\cite{he2022po}, generative post-processing~\cite{hoogeboom2023high}  and strategies to control the trade-off between fidelity and realism~\cite{iwai2021fidelity,agustsson2023multi}. MS-ILLM~\cite{muckley2023improving} introduced a non-binary discriminator conditioned on quantized local image representations, significantly improving statistical fidelity of generative compression. Gao et al.~\cite{10256132} proposed an invertible image generation based framework to improve the quality of the restored images for extremely
low bitrate compression. Recently, probabilistic diffusion models~\cite{ho2020denoising} have drawn increasing attention due to its impressive performance on image generation tasks, with perceptual qualities comparable to GAN based methods while maintaining stable training. Some works use diffusion models for image compression. Emiel et al.\cite{hoogeboom2023high} employ a denoising diffusion model to enhance the quality of reconstructed images. Perco\cite{careil2023towards} proposes using iterative diffusion models for decoding, which improves image quality while eliminating dependency on bitrate. DiffEIC~\cite{DiffEIC} leverages the latent of images in the diffusion space as guidance for compression at extremely low bitrates. However, these diffusion-based approaches are hindered by significant computational costs, substantial memory requirements, and high latency due to their complex designs and multi-iteration inference processes. Additionally, they are prone to introducing artifacts or deviations that may compromise fidelity to the original image.

\begin{figure*}
  \centering
    \includegraphics[width=\textwidth]{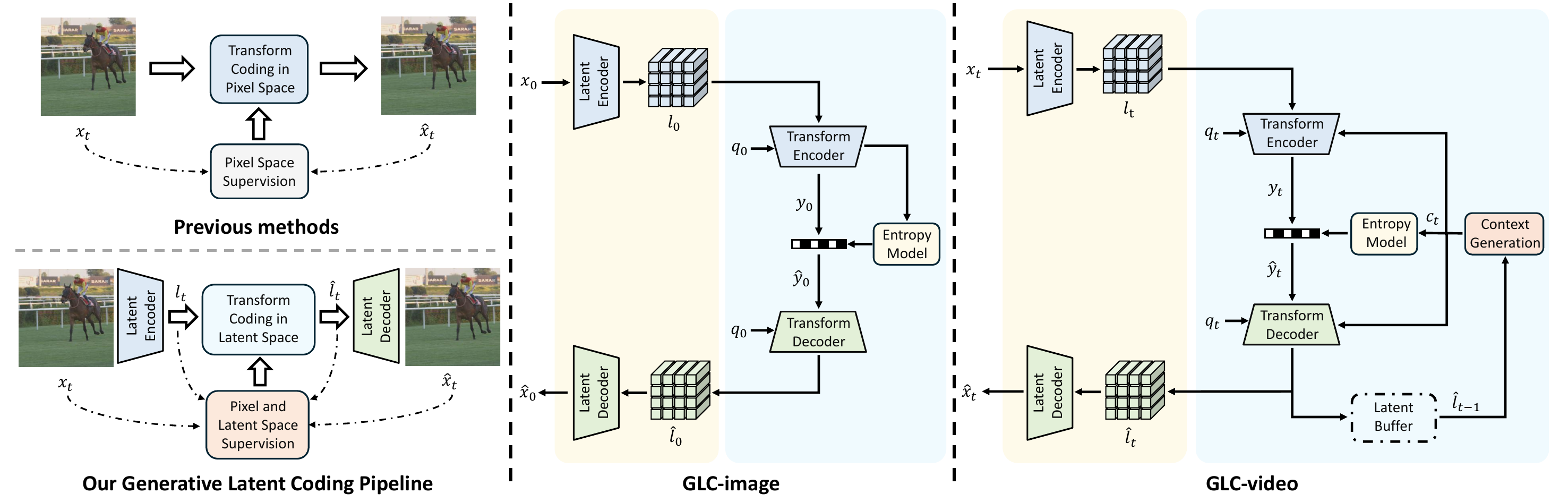}
    \vspace{-5mm}
    \caption{
            Left: Comparison with previous methods. Unlike traditional approaches that perform transform coding in the pixel space, our scheme operates in the generative latent space.
            The generative latent coding pipeline involves three steps: (1) encoding the input into a generative latent space, (2) compressing the latents using transform coding, and (3) decoding the compressed latent to reconstruct the image.
            Center: Illustration of the proposed GLC-image for image compression.
            Right: Illustration of the proposed GLC-video for video compression.}
  \label{fig:Main}
  \vspace{-3mm}
\end{figure*}

\subsection{Learned Video Compression}
Most existing learned video compression methods are designed to optimize traditional metrics like PSNR or MS-SSIM\cite{wang2003multiscale}. 
They can be broadly categorized into two categories: residual coding-based frameworks and conditional coding-based frameworks. The residual coding-based framework~\cite{dvc, lu2020end, fvc, hu2022coarse, ma2024uncertainty} typically compute and compress the residuals. A prediction frame is firstly generated from the previously decoded frame by aligning with the estimated optical flow. Subsequently, the residual between the prediction frame and the current frame is coded for reconstruction. 
The conditional coding-based frameworks do not directly use a subtraction operation to reduce redundancy, but generates temporal context from previously decoded frames or buffered features~\cite{DCVC, canfvc, qi2023motion, li2023neural, qi2024long, DCVC-FM, 10416688}. Compared to residual coding, conditional coding learns temporal context more flexibly. 
Recently, there has been a growing interest in generative video compression, which focuses on improving perceptual quality rather than solely prioritizing PSNR. For instance, PLVC~\cite{yang2022perceptual} incorporates a recurrent model with GAN loss to produce visually appealing reconstructions. Similarly, \cite{li2023high} addresses challenges such as poor reconstruction in newly-emerged regions and checkerboard artifacts caused by deconvolutions and optimization inefficiencies.
In contrast to these methods, our scheme achieves high-quality visual reconstruction at significantly lower bitrates by performing transform coding in the generative latent space, effectively capturing semantic and temporal dynamics for perceptually-driven video compression.

\subsection{Latent Space Modeling}
The latent space modeling technique has been primarily developed for image generation, which involves modeling the distribution of inputs within the latent space of a neural network. Early contributions by Chen et al.~\cite{chen2016variational} and Oord et al.~\cite{van2017neural} introduced the use of PixelCNN~\cite{van2016pixel} within the latent space of VAE and VQ-VAE for image generation. Esser et al.~\cite{esser2021taming} advanced this idea by integrating transformers into the VQ-VAE latent space, enabling high-quality generation. More recently, Rombach et al.~\cite{rombach2022high} achieved remarkable results in high-resolution image generation by employing diffusion models to model the latent space of VQ-VAE.  These studies highlight the potential of utilizing generative latent spaces, particularly the latent space of VQ-VAE.

The concept of latent space modeling has recently been extended to other tasks. For example, CodeFormer~\cite{zhou2022towards} introduced a code prediction transformer that uses distorted latents as input and predicts high-quality VQ-VAE indices for facial restoration. Building on this, Jiang et al.~\cite{jiang2023adaptive} proposed transmitting the predicted indices to facilitate restoration-based facial conferencing. In this work, we explore the latent space modeling for generative image and video compression at ultra-low bitrates. Specifically, we design a transform coding paradigm in the latent space, leveraging its alignment with human perception to deliver superior performance for image and video compression.

\section{Generative Latent Coding}
In this section, we introduce the proposed Generative Latent Coding(GLC) paradigm.
Our approach performs transform coding within the generative latent space, which is more aligned human perception than the pixel space. The latent space is constructed using a Generative Latent Auto-Encoder (detailed in Sec.\ref{generative_latent_auto_encoder}).  Notably, our latent-space transform coding supports variable bitrates within a single model, as discussed in Sec.\ref{rate_variable_latent_transformation}. Additionally, we provide a detailed discussion on the application of latent-space transform coding for image compression (GLC-Image, Sec.\ref{glc-image}) and video compression (GLC-Video, Sec.\ref{glc-video}).

\subsection{Generative Latent Auto-Encoder}
\label{generative_latent_auto_encoder}
To achieve a latent space aligned with human perception, the GLC pipeline employs a generative VQ-VAE~\cite{esser2021taming} as the latent auto-encoder ($E$ and $D$). It maps the input image $x_t$ or video frame $x_t$ into visual semantic elements through a discrete codebook $B$. The codebook size is 16384. Then a generative decoding process is incorporated, ensuring both texture realism and semantic consistency. Additionlly, through training with the discrete codebook $B$ as a variational bottleneck, a low-entropy and distortion-robust latent space is achieved, contributing to the compression process.

The proposed GLC-image and GLC-video share the same structure for the generative latent auto-encoder to ensure the consistency of the latent space, and can be uniformly presented (in the left of Fig.~\ref{fig:Main}). 
The input images or video frames are firstly encoded into the latents using the latent encoder. The latents are then processed by the transform coding, which comprises encoding, scalar-quantization and decoding. Finally, the latent decoder takes the decoded latents as input and generates the reconstructed images or video frames.
The overall process can be formulated as follows:
\begin{equation}
    \begin{aligned}
        l_t = E(x_t),\ & \ y_t = g_a(l_t, c_t) \\
        \hat{y}_t = &\ Q(y_t) \\
        \hat{l}_t = g_s(\hat{y}_t, c_t),\ & \ \hat{x}_t = D(\hat{l}_t)
    \end{aligned}
    \label{equ:generative_latent_auto_encoder}
\end{equation}
Here, $E$ and $D$ denote the latent encoder and decoder, respectively; $g_a$ and $g_s$ represent the analysis transform (implemented using the transform encoder) and the synthesis transform (implemented using the transform decoder), respectively; $x_t$, $l_t$, and $y_t$ correspond to the input image/video frame, the latent, and the code used for entropy coding, respectively; $\hat{\cdot}$ indicates the corresponding reconstruction; $t$ indicates the time step, and $c_t$ denotes the temporal context (for image compression, $t=0$ and $c_t$ is not applicable); and $Q$ stands for scalar quantization.

\subsection{Rate-Variable Latent Transformation}
\label{rate_variable_latent_transformation}
\begin{figure*}
  \centering
  \subfloat[]{
    \includegraphics[width=0.35\linewidth]{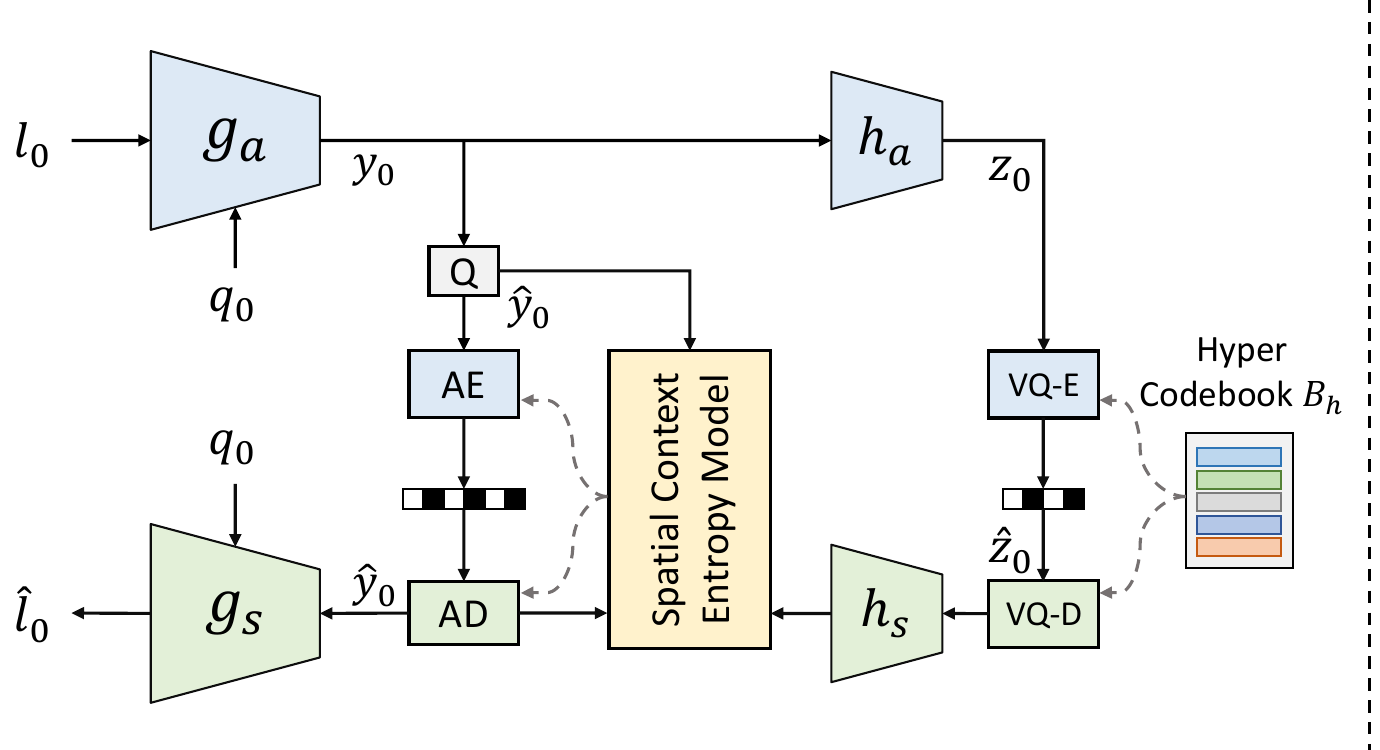}%
    \label{fig:Latent_Compression}
  }
  \hspace{-2mm}
  \subfloat[]{
    \includegraphics[width=0.2\linewidth]{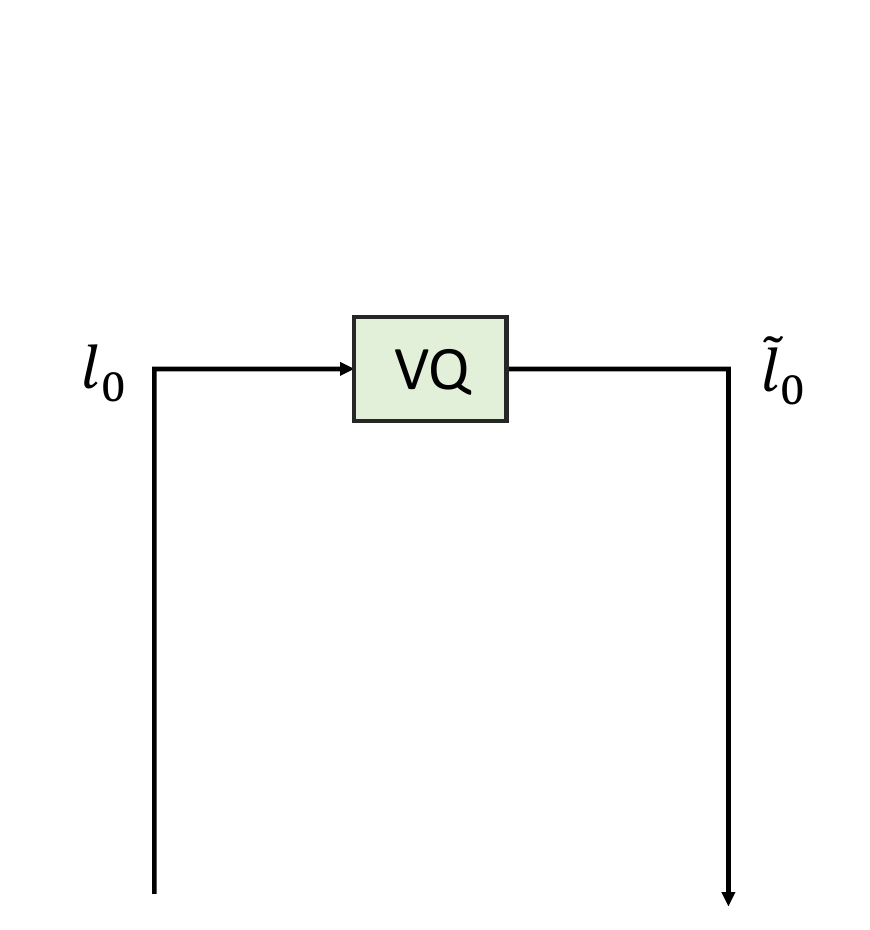}%
    \label{fig:VQ_Hyper_Arch_0}
  }
  \subfloat[]{
    \includegraphics[width=0.2\linewidth]{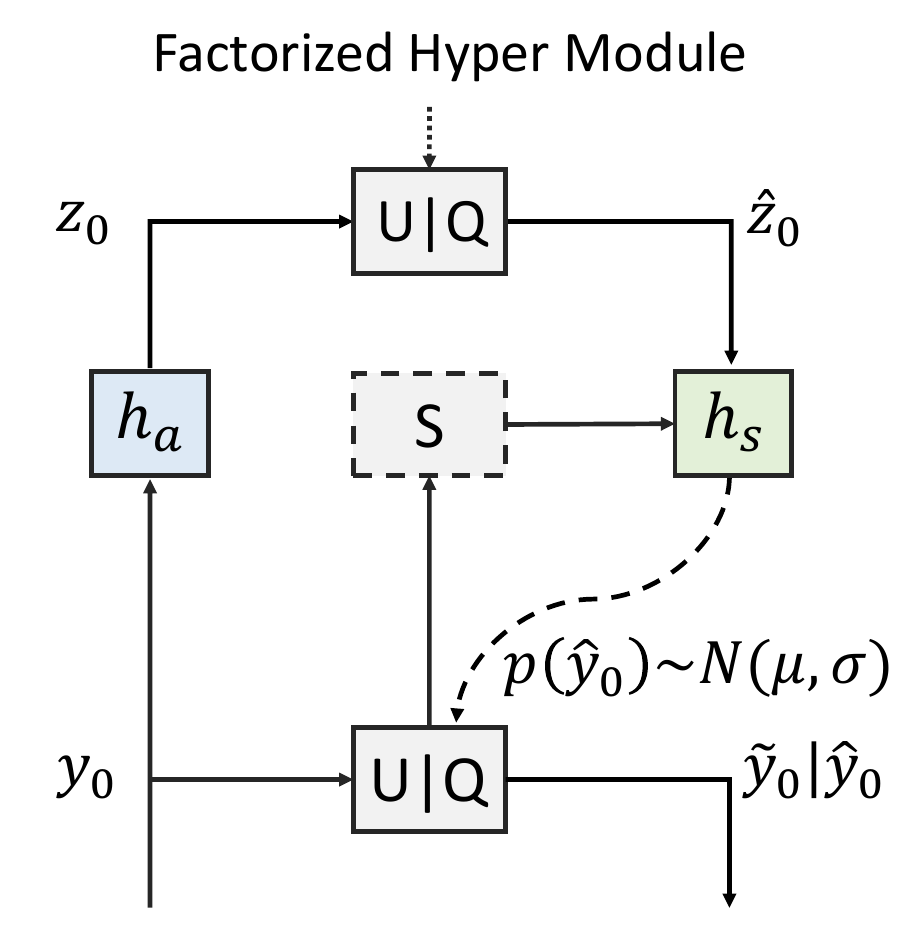}%
    \label{fig:VQ_Hyper_Arch_1}
  }
  \subfloat[]{
    \includegraphics[width=0.2\linewidth]{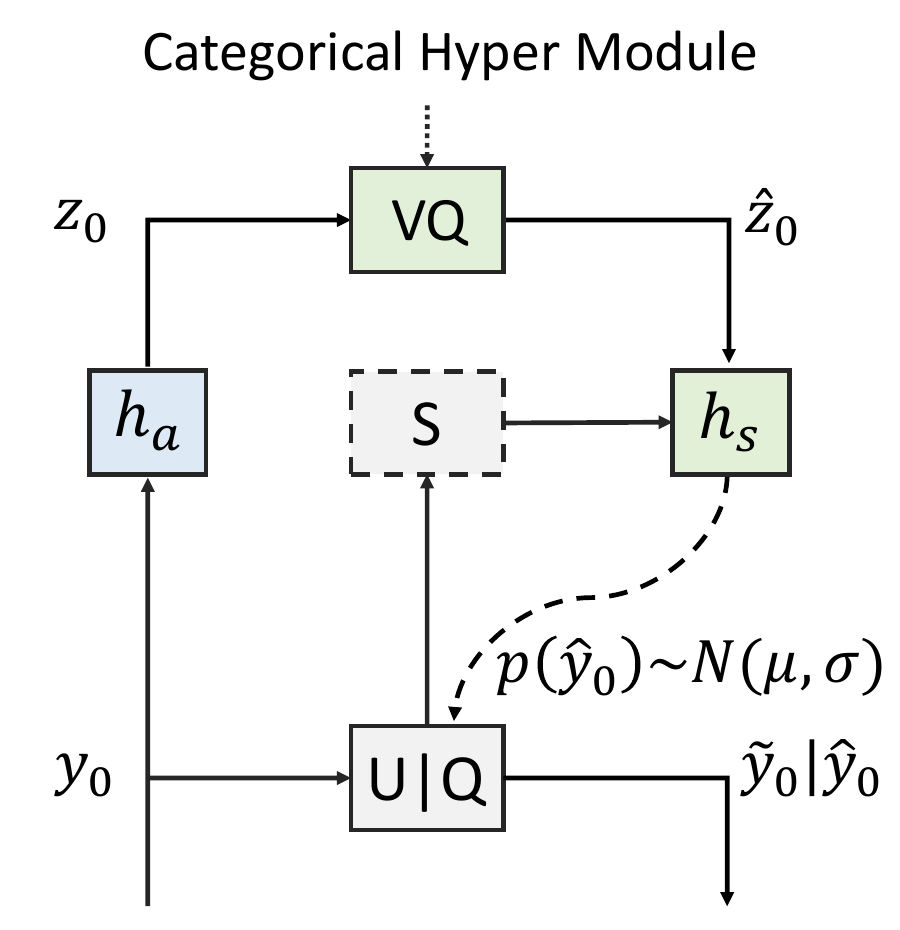}%
    \label{fig:VQ_Hyper_Arch_2}
  }
  \caption{Illustration of the transform coding in the latent space of GLC-image and comparison with other coding schemes in operational diagrams. (a) The model structure of transform coding module in GLC-image. (b) indices-map coding~\cite{mao2023extreme,jiang2023adaptive}, (c) transform coding with factorized hyper module~\cite{balle2018variational} and (d) proposed transform coding with our spatial categorical hyper module. Here, AE and AD denote arithmetic encoding and decoding, VQ-E and VQ-D refer to VQ-indices-map encoding and decoding, Q represents scalar quantization, U signifies the addition of uniform noise as a differential simulation of Q, and S denotes the spatial context entropy module.}
  \label{fig:VQ_Hyper_Arch}
    \vspace{-3mm}
\end{figure*}

A straightforward approach to compressing latents $l_t$ is the VQ-indices-map coding~\cite{jiang2023adaptive, mao2023extreme} (Fig.~\ref{fig:VQ_Hyper_Arch_0}). 
These methods reduce the bit cost of individual feature vectors by encoding them into discrete VQ indices, but they overlook the correlations between different feature vectors, consequently resulting in insufficient redundancy reduction and higher bit cost. To address these limitations, our proposed GLC-image and GLC-video frameworks replace the vector-quantization step with a transform coding module, as depicted in Fig.\ref{fig:Main}. The transform coding module processes latents through advanced network structures, converting them into more compact and efficient forms while preserving essential information, thus effectively reducing redundancy. 
Additionally, for video compression task, transform coding 
facilitates the modeling of inter-frame correlations by flexibly incorporating the temporal context from the previously decoded latents.

An additional advantage of transform coding over VQ-indices-map coding is its inherent support for rate-variable compression, which is crucial for practical image and video codecs. VQ-indices-map coding is constrained by the limited modeling capacity of its codebook, which can only represent one specified distribution. However, different bitrate levels naturally require the modeling of different distributions. By contrast, transform coding maps the latents into a unified Gaussian distribution, with the rate variability achieved through adjustments to the Gaussian parameters, such as the mean and scale. In the proposed GLC pipeline, we incorporate a scaling factor $q_t$ into the transform coding process, following the strategies employed in advanced variable-bitrate neural codecs~\cite{DCVC-HEM, li2023neural, DCVC-FM}. Users can adjust the scaling factor to control the bitrate within a single model, eliminating the need to switch between multiple models.

\begin{figure}
  \centering
    \includegraphics[width=\linewidth]{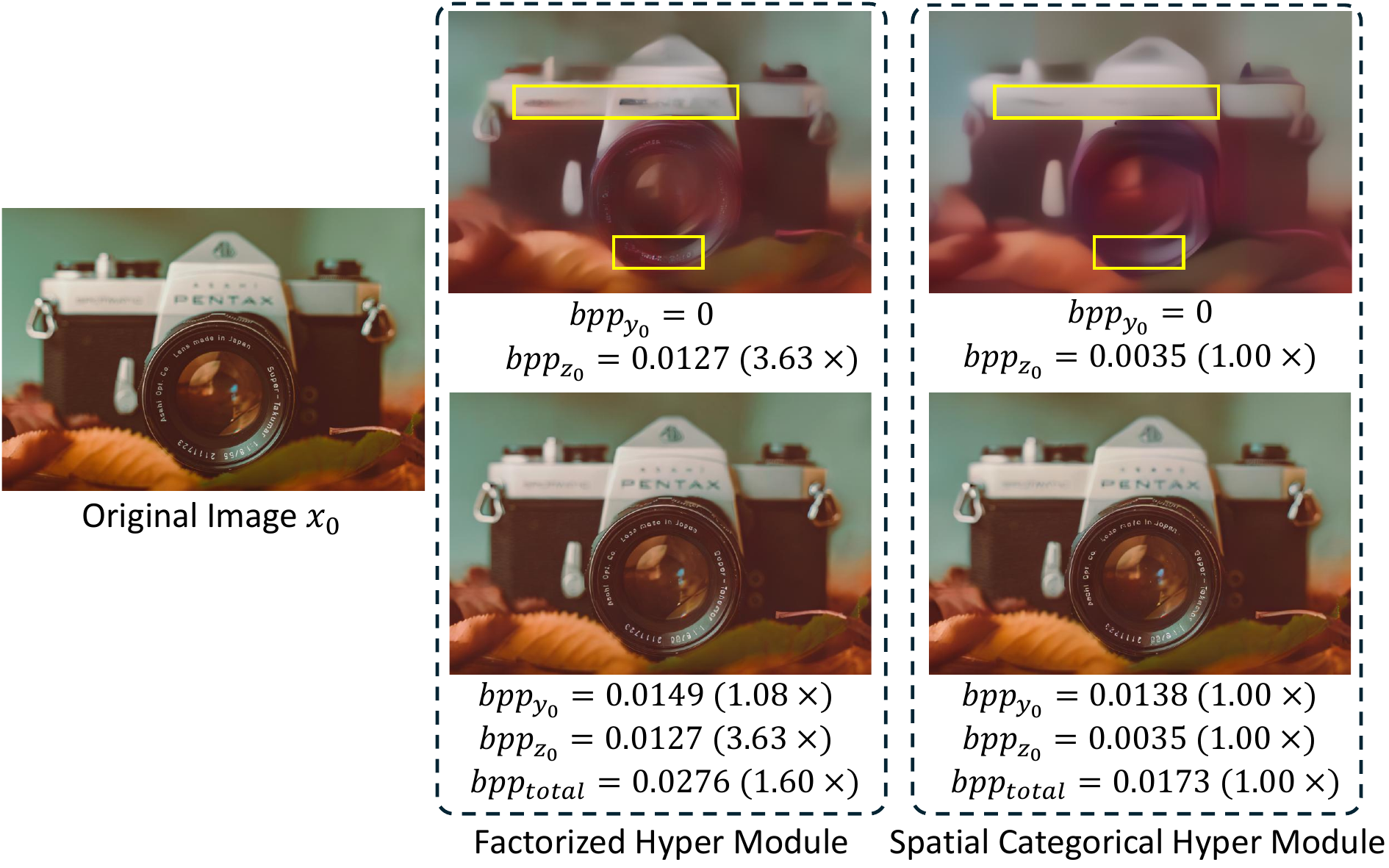}
    \vspace{-6mm}
    \caption{Visual comparison for the spatial categorical hyper module and factorized hyper module. 
    The bit per pixel (bpp) value for coding $y_0$ and $z_0$, and the bpp multiplier relative to our method are shown. The proposed spatial categorical hyper module encodes essential semantic and structural information with significantly fewer bits while being less susceptible to low-level noise, thereby achieving comparable visual quality and substantially reducing the overall bit cost.
    }
  \label{fig:VQ_Hyper_recon_with_z}
    \vspace{-3mm}
\end{figure}

\subsection{Latent-space Transform Coding for GLC-image}
\label{glc-image}
In this section, we detail the design of the latent-space transform coding used in GLC-image. As shown in the center of Fig.~\ref{fig:Main}, the transform coding procedure (Sec.~\ref{glc-image-transform-coding}) consist of a transform encoder, a transform decoder, and an entropy model. Considering the characteristics of hyper information at ultra-low bitrates, we further enhance the coding performance by incorporating a spatial categorical hyper module (Sec.~\ref{categorical_hyper_module}). This module allows for more effective modeling and compression of the hyper information, contributing to the overall efficiency of the system.

\subsubsection{Transform Coding}
\label{glc-image-transform-coding}
The transform encoder and decoder are based on the image codec presented in \cite{li2023neural}, using cascaded depth-wise convolution blocks for efficient image compression. The main transformations include converting latent $l_0$ into the code $y_0$, as well as the inverse process of converting $\hat{y}_0$ back into $\hat{l}_0$. We configure the number of channels of the latent $l_0$ and code $y_0$ as 256. We also incorporate learned scalers $q^{enc}_0$ and $q^{dec}_0$ into the transform encoder and decoder for supporting variable bitrates. The entropy model predicts the probability distribution of the quantized code $\hat{y}_0$ through the proposed spatial categorical hyper module (Section \ref{categorical_hyper_module}) and a spatial context module.
The spatial categorical hyper module utilizes a hyper codebook $B_h$ to code the hyper information $\hat{z}_0$. During inference, the indices are compressed using fixed-length coding, where each code index is encoded into $\log_2{N_b}$ bits, where $N_b$ represents the size of $B_h$. For the spatial context module, we adopt the quantree-partition-based structure from~\cite{li2023neural}, which predicts the probability by leveraging the hyper prior and the previously decoded parts of $\hat{y}_0$.

\begin{figure*}[t]
  \centering
    \includegraphics[width=0.9\linewidth]{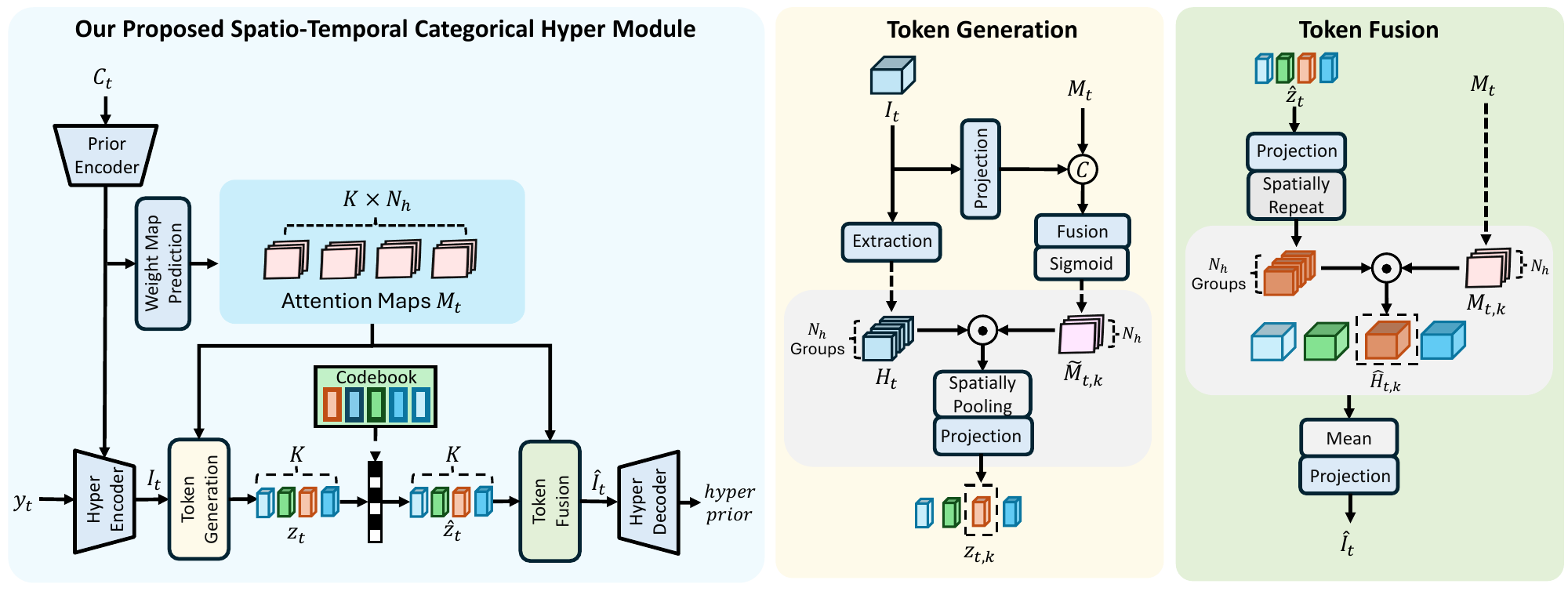}
    \vspace{-3mm}
    \caption{Our proposed spatio-temporal categorical hyper module in video compression. The token generation and token fusion modules are also illustrated.}
  \label{fig:glc-video-hyper-module}
    \vspace{-3mm}
\end{figure*}

\subsubsection{Spatial Categorical Hyper Module}
\label{categorical_hyper_module}

Most image compression frameworks  employ the factorized hyper module~\cite{balle2018variational} (Fig.~\ref{fig:VQ_Hyper_Arch_1}) to model the hyper information $z_0$. We observe that at ultra-low bitrates, the factorized $z_0$ predominantly encodes low-level features, such as color and texture. However, 
it also introduces low-level noise, which not only increase the bit cost of $z_0$, but also necessitates extra bits for correction, resulting in a high bit cost, as highlighted by bounding boxes in Fig.~\ref{fig:VQ_Hyper_recon_with_z}. To address this issue, we propose a spatial categorical hyper module, as shown in Fig.~\ref{fig:VQ_Hyper_Arch_2}. It consists of a hyper analysis transform $h_a$, a hyper synthesis transform $h_s$, and a hyper codebook $B_h$. The hyper codebook is employed to store the basic semantic elements. 
Specifically, the process can be described by the following transformations:
\begin{equation}
    z_0 = h_a(y_0),\ \ \hat{z}_0 = VQ(z_0, B_h), \ \ prior_{z_0} = h_s(\hat{z}_0)
\end{equation}
where $z_0$ represent the hyper-codes, $\hat{z}_0$ refers to the reconstructed hyper-codes, and $VQ(\cdot, B_h)$ denotes vector quantization through nearest lookup in the hyper codebook $B_h$. This codebook is trained to capture basic semantic elements, bringing a better trade-off between preserving essential semantics and reducing the bit cost of coding hyper-information.
As shown in Fig.~\ref{fig:VQ_Hyper_recon_with_z},
when only transmitting the hyper information by $z_0$, both the factorized hyper module and our 
spatial categorical hyper module can enable a rough reconstruction of the input $x_0$. However, the factorized hyper module primarily focuses on encoding low-level details, resulting in significantly higher bit consumption for $z_0$.
In contrast, our spatial categorical hyper module avoids encoding low-level noise and focus on capturing high-level semantic information, preserving essential attributes such as shape and rough color. It is shown that, when transmitting both $y_0$ and $z_0$, the model equipped with the spatial categorical hyper module can achieve comparable visual quality but use much fewer bits than the factorized hyper module, showcasing the effectiveness of our proposed method.

\subsection{Latent-space Transform Coding for GLC-video}
\label{glc-video}
In the preceding section, we discussed the design of GLC-image, which applies transform coding in the latent space for image compression. In this section, we introduce the latent-space transform coding in our GLC-video for video compression. Recognizing that temporal correlations across frames persist in the latent space, GLC-video converts frames into latents and employs conditional coding to reduce the temporal redundancy. As illustrated on the right side of Fig.~\ref{fig:Main}, 
besides the generative latent auto-encoder for converting between the latent space and the pixel space, GLC-video employs a conditional latent codec for latent-space transform coding, which facilitates the coding of the current latent $l_t$ using previous decoded latent $\hat{l}_{t-1}$ as temporal conditions. Particularly, the temporal context is also integrated into the spatio-temporal categorical hyper module (Section~\ref{sec-glc-video-hyper}), enabling it to capture semantic dynamics as hyper information. 

\subsubsection{Conditional Transform Coding}
For the model design, we build our conditional latent codec based on the framework presented in~\cite{DCVC-HEM}. Given that the latent $l_t$ has a resolution of $1/16$ compared to the input $x_t$, we simplify the context generation process by utilizing cascaded depth-wise convolution blocks. These blocks extract single-scale ($1/16$) features as temporal context, replacing the more complex multi-scale approach in~\cite{DCVC-HEM}. Similarly, the transform encoder and decoder are streamlined to integrate only single-scale temporal context. The number of channels for the latent $l_t$, the code $y_t$, and the temporal context is uniformly set to 256. As conditional information, the temporal context is concatenated with the inputs to both the transform encoder and decoder and further assists in entropy modeling. To enhance the efficiency of entropy modeling, we propose a spatio-temporal categorical hyper module. This module incorporates the temporal context into hyper information and employs a hyper codebook to encode the dynamics across adjacent frames into $z_t$. Similar to $z_0$ of GLC-image, $z_t$ is compressed using fixed-length coding. GLC-video is also a rate-variable compression framework, whose bitrate is controlled by the the quantization scaler $q^{enc}_t$ and $q^{dec}_t$ generated according to the quantization parameter $q_t$ from the user input.

\subsubsection{Spatio-Temporal Categorical Hyper Module}\label{sec-glc-video-hyper}

In Section~\ref{categorical_hyper_module}, we demonstrated that the proposed spatial categorical hyper module can effectively capture essential spatial semantics using only a few bits for image compression. 
However, for video compression, it becomes crucial to account for temporal correlations of hyper information and such a module is not sufficiently efficient. 
Firstly, adjacent frames typically share similar semantics, so independently encoding categorical hyper information can introduce significant redundancy. 
Secondly, the dynamics of correlated content in a video often follow regular patterns, but the hyper-features with a local receptive field lacks the capability to capture global and high-level dynamics, such as background movement or camera perspective changes. Unlike the dense hyper information required for image compression, the hyper information in video compression is inherently sparse along the temporal dimension and we can encode it into a more compact representation.
To address these challenges, we propose a spatio-temporal categorical hyper module in GLC-video. It inherits the idea of the spatial categorical hyper module to utilize a discrete codebook to model semantics.
By leveraging temporal context, the proposed hyper module focuses exclusively on capturing changes in semantics between the current and previous frames. Correlated content changes across the entire frame are adaptively aggregated, resulting in a series of tokens $z_t$. Compared to previous hyper features, our representations benefit from the capability of capturing global semantic dynamics, which enhances its compactness and sparsity, leading to a significantly reduced number of tokens, thereby achieving a higher compression ratio.

Fig.~\ref{fig:glc-video-hyper-module} presents the detailed structure of our proposed spatio-temporal categorical hyper module, which operates in three main steps: (1) capturing global semantic dynamics as hyper information into $K$ tokens $z_t$; (2) using a hyper codebook to quantize $z_t$ into $\hat{z}_t$; and (3) restoring feature maps from $\hat{z}_t$ to serve as hyper priors for entropy coding. Specifically, temporal information is extracted from the temporal context $c_t$, which assists in extracting the dynamic information as hyper-information from $y_t$. 
For each token, $N_h$ attention maps are predicted to indicate the regions that the token emphasizes, resulting in a total of $K \times N_h$ maps, which are subsequently utilized in the processes of \textbf{Token Generation} and \textbf{Token Fusion}.

\textbf{Token Generation.} This module selectively encodes the global and high-level dynamics from the entire feature map into the tokens $z_t$. 
To ensure comprehensive representation, $K$ tokens are generated using distinct attention maps, each focusing on specific regions to capture diverse information.
As shown in the center of Fig.~\ref{fig:glc-video-hyper-module}, the generation of each token, $z_{t, k}$ ($k\in\{1, ..., K\}$), is conducted in a group-wise manner. 
To enhance the attention map by incorporating features of the current time step, the attention maps $M_t$ are fused with the hyper information $I_t$, yielding refined attention maps $\tilde{M}_{t,k}$ for the $k$-th token.
Features extracted from the hyper information $I_t$, denoted as $H_{t}$, are divided into $N_h$ groups to reduce the computation complexity of the subsequent operations. Next, weighted pooling is applied to facilitate $z_{t, k}$ to selectively encoding hyper information: $H_{t}$ is firstly multiplied by $\tilde{M}_{t,k}$, (where channels within each group share the same attention map) to emphasize semantic dynamics, followed by global average pooling. Finally, the resulting output is projected onto the token $z_{t, k}$, which will then be coded with the codebook.

\textbf{Token Fusion.} The reconstructed tokens $\hat{z}_t$ encapsulate global and high-level dynamics. To restore spatial-aware information from $\hat{z}_t$, the token fusion process serves as an approximate inversion of Token Generation, as illustrated on the right side of Fig.~\ref{fig:glc-video-hyper-module}. To construct the initial feature maps, we project $\hat{z}_t$ through a linear layer, followed by a spatial repetition process.
To reintroduce the spatial information, we use the predicted attention maps as guidance. Specifically, element-wise multiplication between the $k$-th feature map and the corresponding attention map $M_{t,k}$ is conducted in a group-wise manner. Each attention map is shared among the channels within the corresponding channel group. Finally, to fuse distinct information of the $K$ feature maps, we aggregate them by computing their average values, and the output is refined using convolutional layers to synthesize the spatially-aware hyper information $\hat{I}_t$. This restored hyper information is then fed into the hyper decoder to generate the hyper prior, which assists entropy coding.

\begin{figure*}[t]
  \centering
    \includegraphics[width=0.95\textwidth]{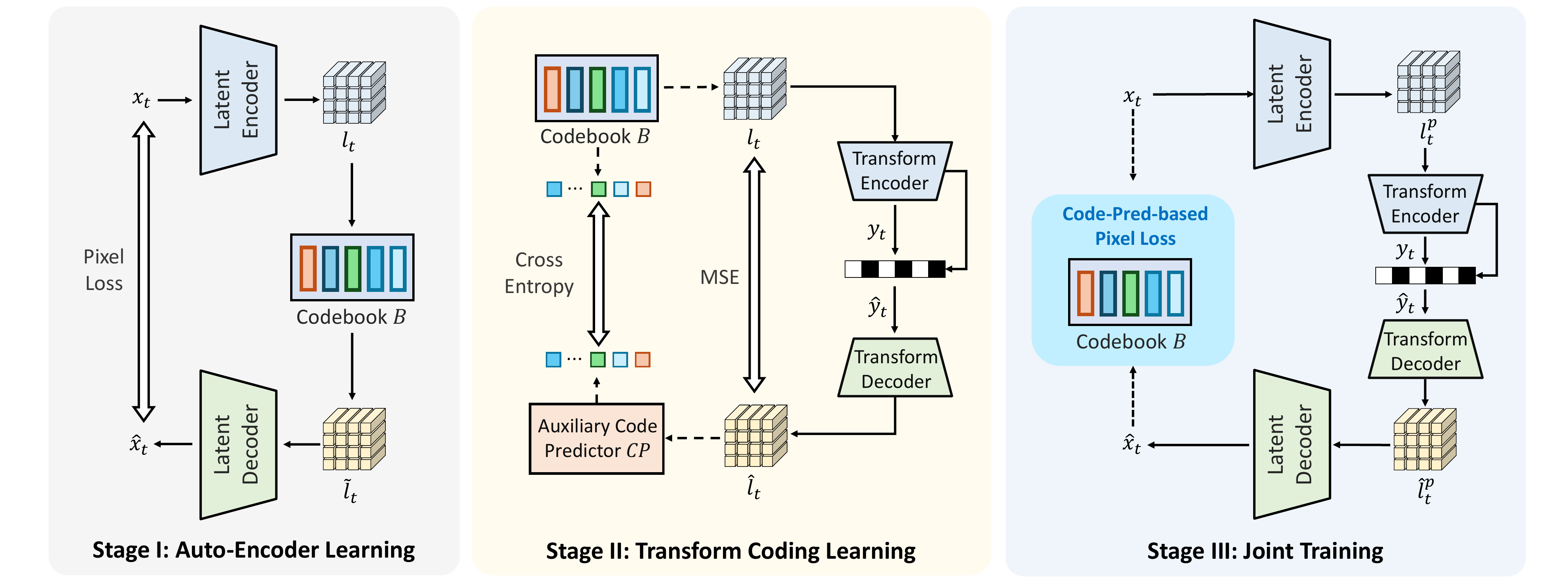}
    \caption{The progressive training procedure consists of three stages (see Section \ref{progressive_training}). Stage I: Train the generative VQ-VAE to establish a latent space aligned with human perception. Stage II: Train the transform coding module to effectively compress latents, guided by code-prediction-based latent supervision. Stage III: Jointly fine-tune the entire network using code-prediction-based pixel supervision to further enhance performance.}
  \label{fig:progressive_training}
    \vspace{-3mm}
\end{figure*}

\section{Progressive Training of GLC}
\label{progressive_training}
In this section, we detail a progressive training strategy designed to fully leverage the potential of our generative latent coding pipeline, as illustrated in Fig.~\ref{fig:progressive_training}. This strategy is structured into three distinct stages, and each stage employs customized loss functions to guide the training of different components, ensuring that each component of the model contributes effectively to the overall performance: 1. Stage I: Auto-Encoder Learning (Section~\ref{sec:stage1}), where the focus is on learning a perceptually aligned latent space to ensure high-quality reconstructions. 2. Stage II: Transform Coding Learning (Section~\ref{sec:stage2}), which involves training the model to perform efficient transform coding on the learned latent space, targeting ultra-low bitrates. 3. Stage III: Joint Training (Section~\ref{sec:stage3}), which finetunes the entire network to optimize the overall compression performance. 

\subsection{Stage I : Auto-Encoder Learning}
\label{sec:stage1}
The primary goal of this stage is to establish a latent space that aligns with human perception, improving the quality of reconstructed images. We trains a generative VQ-VAE as the initialization for the latent encoder $E$ and decoder $D$.  To promote sparsity in the latent space, an auxiliary codebook $B$ is introduced for nearest vector-quantization, transforming the latent $l_t$ into $\widetilde{l}_t$. The training objective $\mathcal{L}_{AE}$ combines multiple loss functions: reconstruction loss, perceptual loss~\cite{johnson2016perceptual}, adversarial loss~\cite{goodfellow2014generative}, and codebook loss~\cite{van2017neural}:

\begin{equation}
    \begin{aligned}
        \mathcal{L}_{AE} &= ||x_t-\hat{x}_t|| + \mathcal{L}_{lpips}(x_t, \hat{x}_t) + w_{adv}\cdot\mathcal{L}_{adv}(x_t, \hat{x}_t) 
        \\
        &+ \underbrace{||\text{sg}(l_t)-\widetilde{l}_t|| + \beta \cdot ||\text{sg}(\widetilde{l}_t)-l_t||}_{\mathcal{L}_{codebook}}
    \end{aligned}
    \label{equ:codebook_loss}
\end{equation}

Here, $\mathcal{L}_{lpips}$ represents the perceptual loss computed using LPIPS, which leverages VGG features~\cite{simonyan2014very} to measure perceptual similarity. Additionally, $w_{adv} = 0.8$ is the weight assigned to the adaptive Patch-GAN adversarial loss~\cite{esser2021taming}, promoting realism in the reconstructed images. $\text{sg}(\cdot)$ denotes the stop-gradient operator and $\beta=0.25$ is utilized to control the update rates of the $E$ and $B$.

\subsection{Stage II : Transform Coding Learning}
\label{sec:stage2}
With the perceptually aligned latent space established, the next step involves training the transform coding module for compressing latents at ultra-low bitrates. During this stage, the auto-encoder ($E$ and $D$) is fixed. To improve semantic consistency, we introduce an auxiliary code predictor ($CP$), which consists of 9 tansformer layers~\cite{vaswani2017attention} and a linear classifier. It necessitates the latent to possess the capability to predict the correct VQ-indices.

As shown in Fig.~\ref{fig:progressive_training}, the latent $l_t$ is encoded into VQ-indices by $V_{l_t} = VQ(l_t, C)$. Subsequently, these indices are predicted by $\hat{V}_{\hat{l}_t} = CP(\hat{l}_t)$. The code-prediction-based loss can be formulated by
\begin{equation}
    \mathcal{D}_{code}(l_t, \hat{l}_t)=\alpha\cdot CE(V_{l_t}, \hat{V}_{\hat{l}_t}) + ||l_t-\hat{l}_t||_2^2
    \label{equ:code_pred_based_loss}
\end{equation}
where $CE$ denotes the cross entropy loss and $\alpha$ is set to $0.5$ by default. The transform coding module is supervised with the following rate-distortion objective:
\begin{equation}
    \mathcal{L}_{\text{TC}}=\mathbf{E}_{x_t\sim p_{X_t}}[\mathcal{R}(\hat{y}_t)+\lambda\cdot\mathcal{D}_{code}(l_t, \hat{l}_t)]
\end{equation}
where $\mathcal{R}$ represents the estimated rate, and $\lambda$ is used to the control the trade-off between rate and distortion. It is worth noting that the hyper codebook $B_h$ in the spatial categorical hyper module requires to be trained by the codebook loss (as formulated in Equation \ref{equ:codebook_loss}). For conciseness, it is omitted in the loss functions of stage II and stage III. 

\subsection{Stage III : Joint Training}
\label{sec:stage3}
In the final stage, the entire network undergoes joint fine-tuning to further improve compression performance. The code-prediction-based latent supervision is extended into the pixel space and the training procedure is guided with the pixel space supervision. As illustrated in Fig.~\ref{fig:progressive_training}, the encoder $E_{VQ}$, which is trained from stage I, is utilized to encode $x_t$ and $\hat{x}_t$ into latent space by $\hat{l}^{p}_t = E_{VQ}(\hat{x}_t)$ and $l^{p}_t = E_{VQ}(x_t)$. And the code-prediction-based pixel supervision can be calculated by $\mathcal{D}_{code}(l^p_t, \hat{l}^p_t)$ in the same formulation with Equation \ref{equ:code_pred_based_loss}. Here $E_{VQ}$ is used because it can map the input to a compatible latent space with the codebook $B$ for code prediction. The pixel space supervision can be described as : 
\begin{equation}
    \begin{aligned}
        \mathcal{D}_{\text{JT}}&= ||x_t-\hat{x}_t|| + \mathcal{L}_{lpips}(x_t, \hat{x}_t) \\ 
        &+ \lambda_{adv}\cdot\mathcal{L}_{adv}(x_t, \hat{x}_t) + \lambda_{code}\cdot\mathcal{D}_{code}(l^p_t, \hat{l}^p_t) \\
    \end{aligned}
\end{equation}
where we set $\lambda_{code}=0.05$ by default. The overall objective becomes: 
\begin{equation}
    \mathcal{L}_{\text{JT}}=\mathbf{E}_{x_t\sim p_{X_t}}[\mathcal{R}(\hat{y}_t)+\lambda\cdot\mathcal{D}_{\text{JT}}]
\end{equation}

\subsection{Discussion of Code-Prediction-Based Loss}

Recent studies~\cite{zhou2022towards, jiang2023adaptive} highlight the effectiveness of code-prediction transformers for high-quality reconstructions, which typically feed the predicted latent directly into the decoder for reconstruction. Different from these methods, our approach considers code prediction solely as an auxiliary supervision during training, without incorporating it into the inference process of the compression pipeline.

This design is grounded in the following observation: when a code prediction module is introduced before the decoder, the fine-tuning process in stage III cannot further enhance compression performance. It arises because the codebook becomes a performance bottleneck, restricting the input of the decoder to the vector-quantized latent, which has already been well-trained in stage I. In our approach, this bottleneck is eliminated by utilizing the code prediction solely as auxiliary supervision during training. Specifically, the VQ operation with codebook $B$ is utilized only during training stage I (auto-encoder learning) of progressive training. In the subsequent stages, it serves merely as an auxiliary supervision and no longer strictly constrains the information of $l_t$. Consequently, in stage II, stage III, and during inference, the codebook $B$ is not required for coding $l_t$ (only the hyper VQ codebook $B_h$ is necessary for the hyper module). It allows more flexible input for the decoder, benefiting the additional fine-tuning. This code-prediction-based supervision can effectively improve the semantic consistency of the reconstructed images. By necessitating the latent to have the ability for predicting the code index, the latent inherently encodes richer semantic information, such as gestures and attributes.

\begin{figure}[t]
  \centering
    \includegraphics[width=\linewidth]{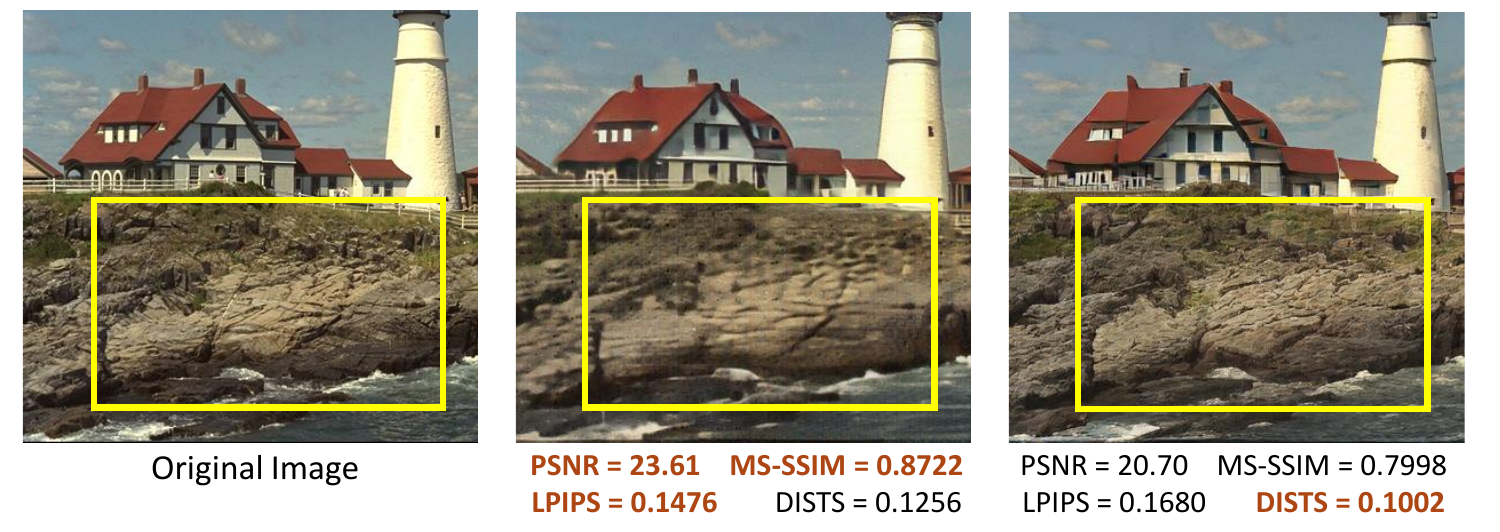}
    \vspace{-8mm}
    \caption{A visual example comparing pixel-level metrics (PSNR, MS-SSIM and LPIPS) and image-level metric DISTS.}
  \label{fig:LPIPS}
  \vspace{-3mm}
\end{figure}

\section{Experiments}
\subsection{Implementation Details}

\textbf{Structure of Latent Auto-Encoder.}
Considering the generative capability and sparse latent space, we employ the generative VQVAE model~\cite{esser2021taming} as the latent auto-encoder. We adopt the same structure as VQGAN~\cite{esser2021taming}, with a latent resolution of $f =1/16$ of the original input and a codebook size of 16384.

\textbf{Training details}. 
For GLC-image, we also follow \cite{jia2024generative} and use OpenImage test set~\cite{kuznetsova2020open} for training, using randomly cropped $256\times256$ patches. For GLC-video, we follow the strategy adopted by \cite{DCVC-FM}, where $256\times256$ patchs are randomly cropped from vimeo dataset~\cite{vimeo} for training.
Both models are optimized by AdamW~\cite{loshchilov2018fixing} with a batch size of 8. For each batch, we train the model with different $\lambda$ to achieve rate-variable compression ($\lambda\in[0.08 , 0.32]$ for GLC-image and $\lambda\in[0.12, 1.6]$ for GLC-video).\\
\textbf{Evaluation dataset}. We evaluate GLC-image on CLIC 2020 test set~\cite{toderici2020clic} with original resolution for image compression. We also show the results on Kodak~\cite{kodak}, DIV2K~\cite{agustsson2017ntire} and MS-COCO 30K \cite{lin2014microsoft}. For evaluating GLC-video, we employ 1080p datasets, including UVG~\cite{uvg}, MCL-JCV~\cite{mcl-jcv} and HEVC Class B~\cite{sullivan2012overview}.

\textbf{Evaluation metrics}. For image compression, visual quality is evaluated using the reference perceptual metrics DISTS~\cite{ding2020image} and LPIPS~\cite{johnson2016perceptual}, along with the no-reference perceptual metrics FID~\cite{heusel2017gans} and KID~\cite{binkowski2018demystifying}. For video compression, visual quality is assessed using DISTS and LPIPS. In both image and video compression, bits per pixel (bpp) is used to quantify the bitstream size.

\begin{figure*}
  \centering
    \includegraphics[width=\linewidth]{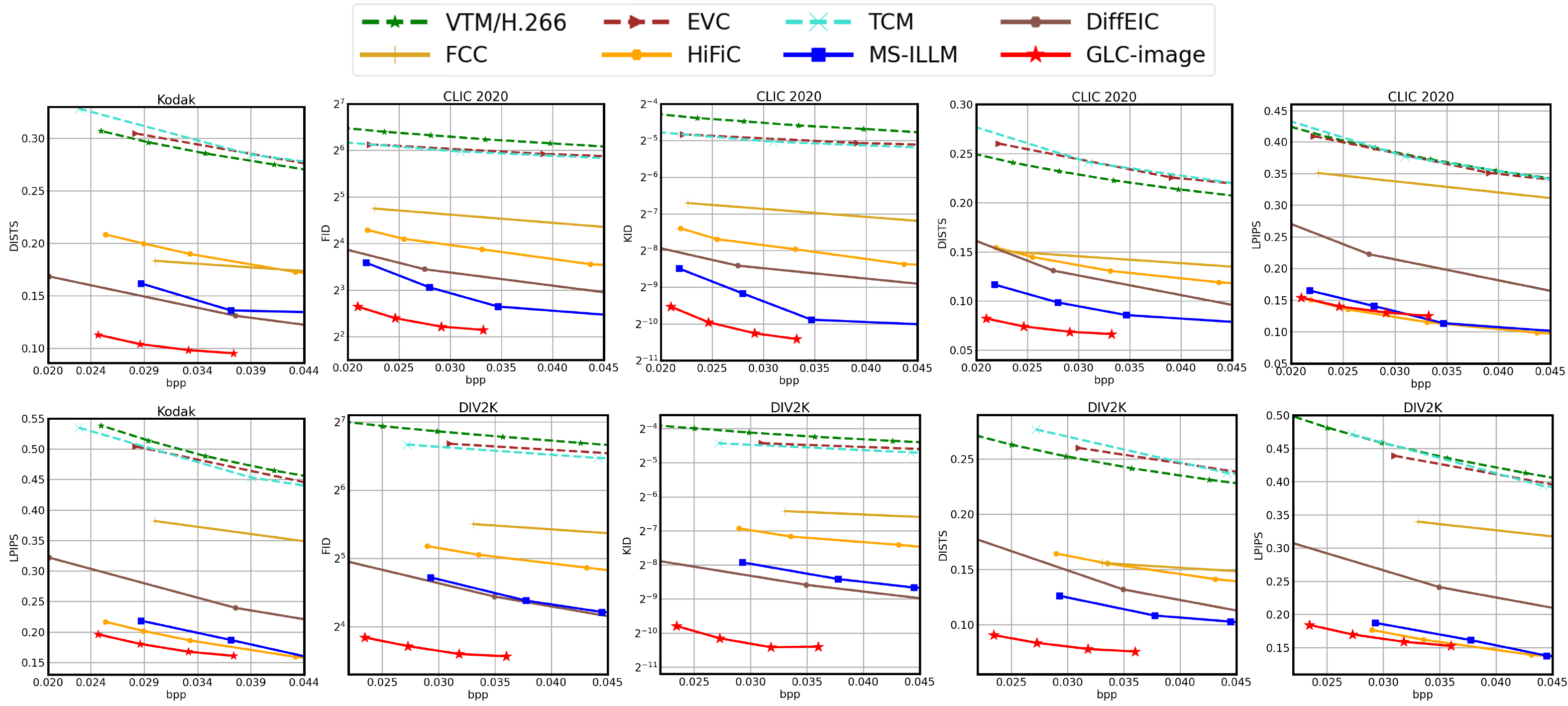}
    \vspace{-5mm}
    \caption{Rate-Distortion curves for comparing DISTS, FID, KID and LPIPS of the proposed GLC-image and other methods on Kodak, CLIC 2020 test set and DIV2K datasets.}
  \label{fig:RD}
  \vspace{-3mm}
\end{figure*}

\begin{figure}
  \centering
    \includegraphics[width=0.6\linewidth]{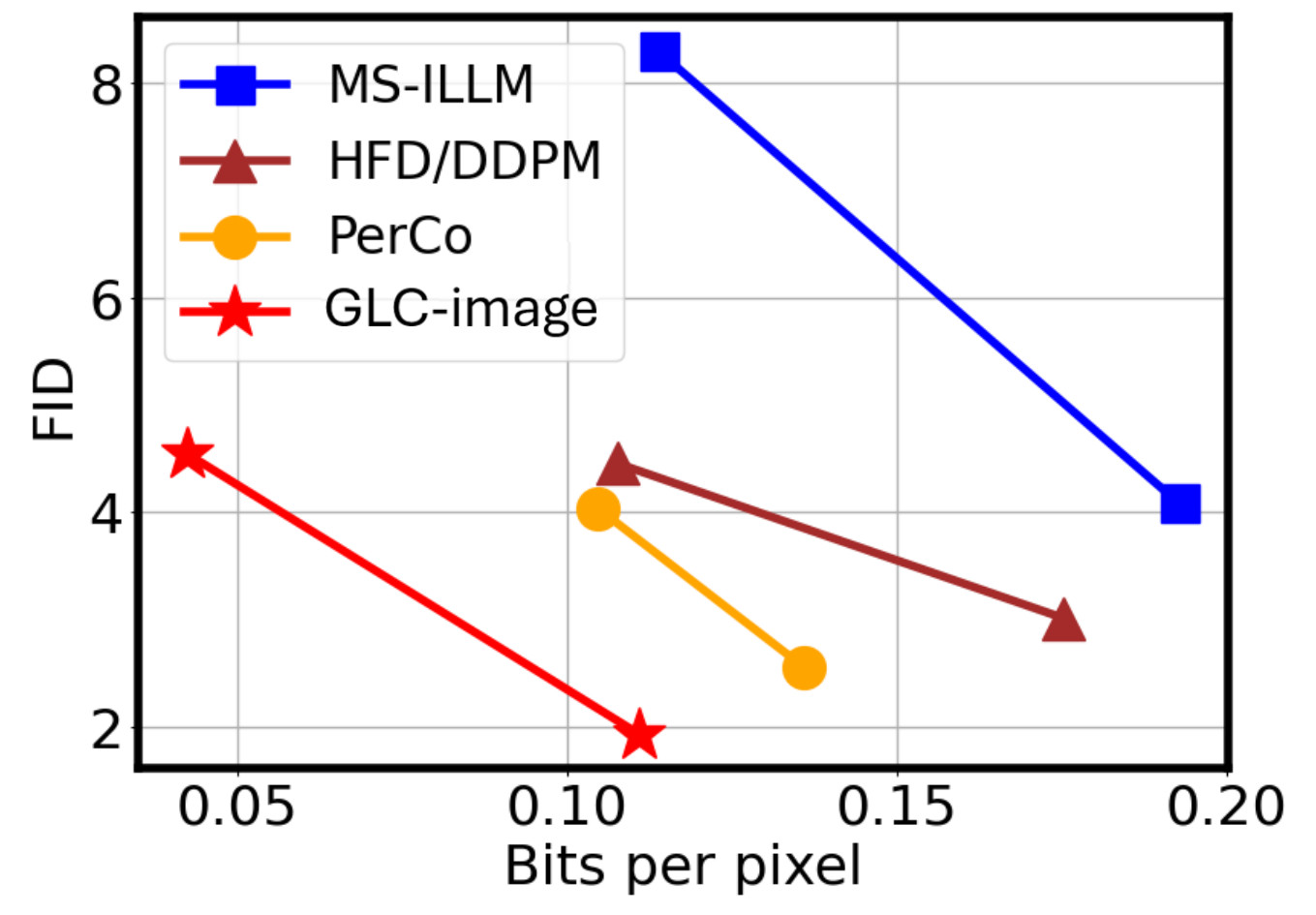}
    \vspace{-3mm}
    \caption{Comparison with recent methods on MS-COCO 30K dataset.}
  \label{fig:ms_coco_30k}
\end{figure}

\begin{figure*}
  \centering
    \includegraphics[width=0.6\linewidth]{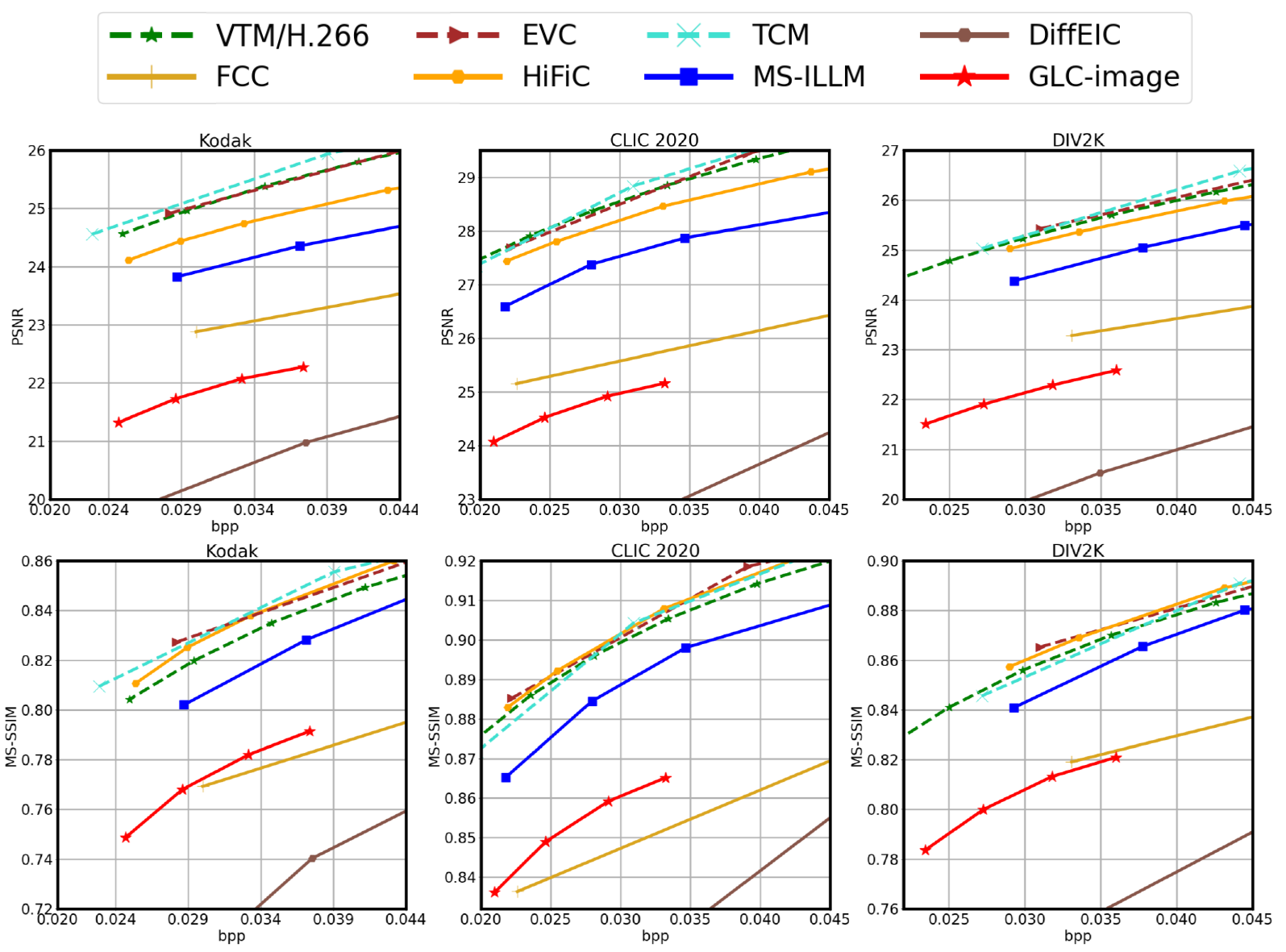}
    \vspace{-3mm}
    \caption{Rate-Distortion curves for comparing PSNR and MS-SSIM of the proposed GLC-image and other methods on Kodak, CLIC 2020 test set and DIV2K datasets.}
  \label{fig:glc-image-rd-psnr}
  \vspace{-3mm}
\end{figure*}

\textbf{Discussion about Measurements}. It is worth noting that the pixel-level distortion metric LPIPS has inherent limitations for assessing visual quality, particularly when evaluating compression methods at ultra-low bitrates, as discussed in~\cite{lei2023text+, ding2020image}. These limitations stem from the fact that LPIPS prioritizes on pixel-level accuracy rather than semantic consistency and texture realism. As illustrated in Fig.~\ref{fig:LPIPS}, the superior result is highlighted in brown. The image on the right is perceptually superior to the one in the middle, despite having lower PSNR, MS-SSIM, and LPIPS scores. In contrast, the image-level metric DISTS offers a more accurate evaluation of perceptual quality. Consequently, this paper primarily focuses on DISTS, FID, and KID as evaluation metrics, rather than LPIPS.

\textbf{Computation of FID and KID}. For image compression, following established practices in generative image compression methods~\cite{mentzer2020high, muckley2023improving}, evaluation is conducted by splitting each image into $256\times256$ patches. Specifically, an $H\times W$ image is divided into $\lfloor H/256 \rfloor \cdot \lfloor W/256 \rfloor$ patches. The extraction origin is then shifted by $128$ pixels in both dimensions to generate an additional set of $(\lfloor H/256 \rfloor - 1) \cdot (\lfloor W/256 \rfloor - 1)$ patches. 
Consistent with~\cite{mentzer2020high, muckley2023improving}, FID and KID are not reported for Kodak~\cite{kodak} due to the limited number of patches (192) derived from its 24 images. 

\textbf{Baseline methods}. For image compression, we compare with traditional codec VTM/H.266~\cite{bross2021overview}, neural image codec TCM~\cite{liu2023learned}, EVC~\cite{guo2022evc}, and generative image codec FCC~\cite{iwai2021fidelity}, HiFiC~\cite{mentzer2020high}, MS-ILLM~\cite{muckley2023improving}. As some methods do not release models for ultra-low bitrate, we either retrain or fine-tune their models to suit such low bitrate. In addition, we also compare with recent difussion based methods~\cite{DiffEIC, careil2023towards, hoogeboom2023high}. 
For video compression, we compare against traditional codecs HM/H.266 and H.266/VTM~\cite{bross2021overview}, previous advanced neural video codec DCVC-FM~\cite{DCVC-FM}, and generative video codec PLVC~\cite{yang2022perceptual}. 
For each video sequence, we evaluate 96 frames where the intra-period is set to -1.

\subsection{Experiments for Image Compression}
In this section, we present the primary results of our GLC-image framework on widely used benchmark datasets and perform ablation studies to validate the effectiveness of each proposed component. For the ablation studies, we evaluate the BD-Rate~\cite{bjontegaard2001calculation} based on the FID-BPP curve using the CLIC 2020 test set.

\textbf{Performance Comparisons.} Fig.~\ref{fig:RD} shows the performance of the GLC-image and compared methods at ultra-low bitrate. It is noted that we utilized an 80G NVIDIA A100 Tensor Core GPU, ensuring that DiffEIC~\cite{DiffEIC} had sufficient memory to be tested at the original resolution. On CLIC 2020, GLC-image demonstrates superiority in terms of DISTS, FID and KID than other methods. Specifically, GLC-image saves about 45\% bits compared to MS-ILLM while maintaining an equivalent FID. When comparing the pixel-level metric LPIPS, GLC-image also achieves comparable performance with high-fidelity generative codecs such as HiFiC and MS-ILLM. In addition, we compare the proposed GLC-image with recent works HFD \cite{hoogeboom2023high} and PerCo \cite{careil2023towards}, along with MS-ILLM, on the MS-COCO 30K dataset \cite{lin2014microsoft}. Following the methodology of \cite{hoogeboom2023high}, we select the same images as them from the 2014 validation set to generate $256\times256$ patches. To match the quality range of their models, we further train a codec around $0.12$ bpp for comparison (the correspondencing latent auto-encoder has $f=\frac{1}{8}$ and $M=256$). As shown in Fig.~\ref{fig:ms_coco_30k}, our model exhibits significant performance improvement.

\begin{figure*}
  \centering
\includegraphics[width=\linewidth]{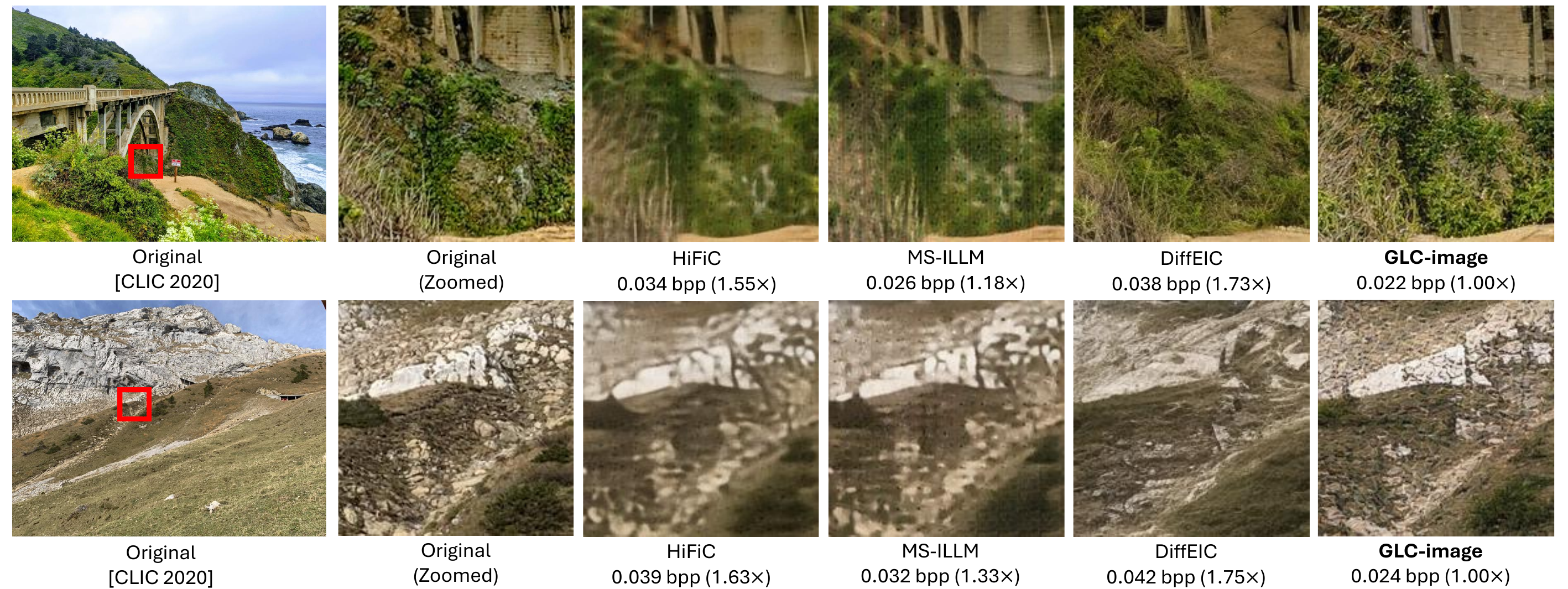}
    \vspace{-5mm}
    \caption{Qualitative examples for comparing the proposed GLC-image with other methods. For each method, we not only present the bit per pixel (bpp) value for coding the image, but also show the bpp multiplier relative to our method.}
  \label{fig:Compare}
  \vspace{-3mm}
\end{figure*}

Fig.~\ref{fig:glc-image-rd-psnr} presents the Rate-Distortion (R-D) curves for PSNR and MS-SSIM on GLC-image. While our model underperforms non-generative methods in terms of PSNR and MS-SSIM. This is expected, as it prioritizes perceptual quality enhancement in ultra-low-bitrate scenarios, where generated details may reduce these metrics. Notably, unlike diffusion-based approaches such as DiffEIC~\cite{DiffEIC}, our method avoids severe semantic distortions, as demonstrated in Fig.~\ref{fig:Compare}.

\begin{figure}
  \centering
    \includegraphics[width=\linewidth]{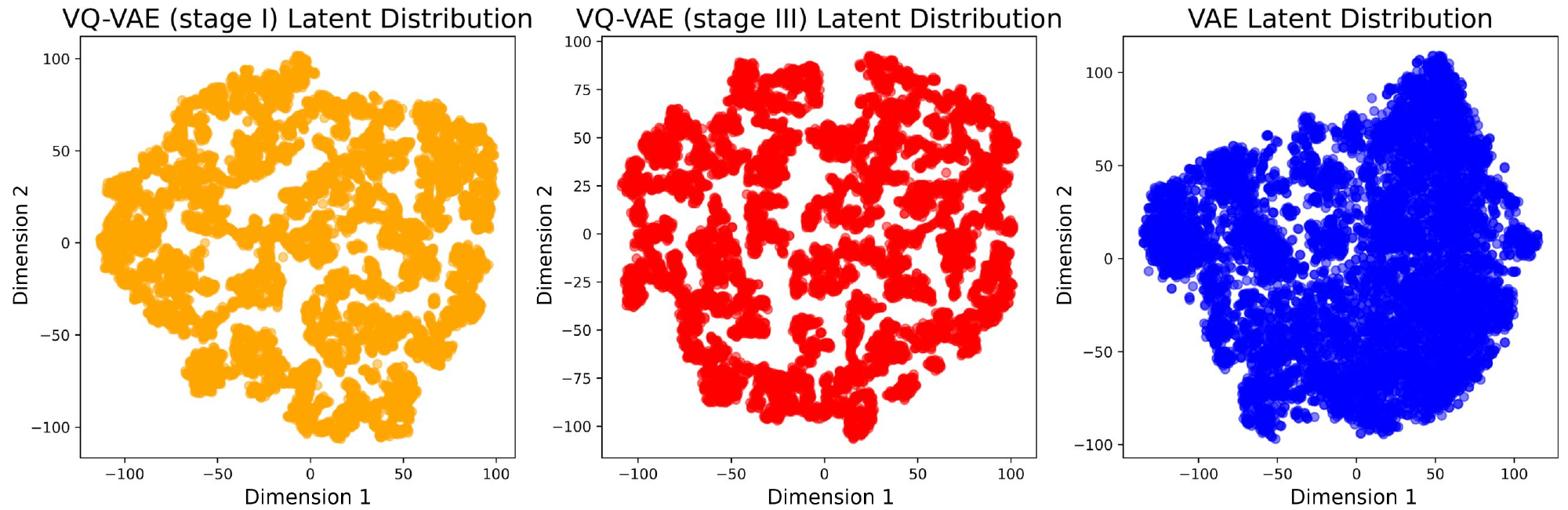}
    \vspace{-3mm}
    \caption{Visulization by t-SNE for comparing the latent distribution on Kodak~\cite{kodak} dataset.}
  \label{fig:glc-image-t-sne}
  \vspace{-3mm}
\end{figure}

\textbf{Visual Comparisons.} Fig.~\ref{fig:Compare} illustrates the qualitative comparison results. Existing methods struggle to produce satisfactory reconstructions at ultra-low bitrates, either suffering from blurry artifacts or yielding low-fidelity outputs. For instance, DiffEIC often fails to retain essential semantic details, such as the surging waves in the first example and the texture on the surface in the second example. TCM, HiFiC, and MS-ILLM generate blurry reconstructions, losing crucial details in the highlighted regions. In contrast, GLC-image delivers visually appealing results, excelling in both realism and fidelity, even at a lower bitrate.

To assess the impact of training stage III (joint training) on latent space sparsity, Fig.\ref{fig:glc-image-t-sne} shows a t-SNE\cite{van2008visualizing} visualization of latent distributions on the Kodak~\cite{kodak} dataset. Note that vector quantization is not applied to the latents in either stage I or III. These results reveal that VQ-VAE trained by stage I (auto-encoder learning) produces sparser and more clustered latents than VAE. Importantly, the latent encoder in stage III (joint training) preserves this sparsity, despite the absence of vector quantization by codebook $B$. In our method, vector quantization on $l_t$ is only used during stage I as an information bottleneck. These findings suggest that our model can benefit from the sparser and more structured latent space of VQ-VAE, which contribute to improved coding efficiency and robustness at extremely low bitrates.

\textbf{Ablation Study on transform coding}. A straightforward approach to compress the VQ-VAE latents is indices-map coding~\cite{mao2023extreme, jiang2023adaptive}. However, it causes $66.2\%$ performance loss compared with transform coding, as shown in Table \ref{tab:transform_coding}. It shows the effectiveness of transform coding on reducing redundancy.

\textbf{Ablation Study on spatial categorical hyper module}. In Section \ref{categorical_hyper_module}, we highlight the advantages of employing our spatial categorical hyper module to model $z$ compared to the commonly used factorized based hyper module. Table \ref{tab:transform_coding} further provides a quantitative comparison, demonstrating a significant improvement of $17.7\%$ with this design.

\begin{table}[t]
    \caption{Ablation study on latent-space compression for GLC-image.}
    \centering
    \resizebox{0.96\linewidth}{!}{
	\begin{tabular}{c | c | c}
	    \midrule
		  Latent coding scheme & Hyper module & BD-Rate $\downarrow$ \\
	    \midrule
  	    Indices-map coding   & -                 & 66.2\%\\
	    \midrule
		\multirow{2}{*}{\textbf{Transform coding}}      & Factorized based  & 17.7\%\\
	    ~                                      & \textbf{Categorical based} & \textbf{0\%} \\
	    \midrule
	\end{tabular}
	}
  \label{tab:transform_coding}
\end{table}

\begin{table}[t]
    \caption{Ablation study on the code prediction module.}
    \centering
    \resizebox{0.85\linewidth}{!}{
        \tabcolsep=10pt
	\begin{tabular}{l | c c }
	    \midrule
		  Code prediction usage & BD-Rate $\downarrow$\\
	    \midrule
		  w/o code pred.   &  13.1\%\\
		  code pred. in network   & 60.7\%\\
		  \textbf{code pred. as supervision} & \textbf{0\%}\\
	    \midrule
	\end{tabular}
	}
  \label{tab:code_prediction}
    \vspace{-5mm}
\end{table}

\textbf{Ablation Study on code-prediction-based supervision}. Section \ref{progressive_training} introduces the use of a code-prediction module as an auxiliary loss during training, instead of during the inference process of the model pipeline as in prior works~\cite{zhou2022towards, jiang2023adaptive}. As shown in Table \ref{tab:code_prediction}, incorporating the code prediction module directly into the network leads to a drastic $60.7\%$ performance drop.
Furthermore, we conducted an additional comparison by removing the code-prediction-based supervision. The results show that adopting the code-prediction-based supervision yields a $13.1\%$ performance improvement.

\begin{table}[tb]
    \caption{Complexity comparison for GLC-image on Kodak with resolution of 512 $\times$ 768. }
    \centering
    \resizebox{0.9\linewidth}{!}{
	\begin{tabular}{c | c c | c | c}
	    \midrule
		  \multirow{2}{*}{Model} & \multicolumn{2}{c|}{Latency (ms)} & \multirow{2}{*}{Params} & \multirow{2}{*}{BD-Rate$\downarrow$} \\
		  ~ & Enc. & Dec. & ~ &  ~ \\
	    \midrule
		  MS-ILLM & 41.8 & 53.5 & 181 M & 0 \\
		  GLC-image    & 37.1 & 58.6 & 105 M & -68.5\%\\
	    \midrule
	\end{tabular}
	}
  \label{tab:complexity_natural}
    \vspace{3mm}
  \caption{Module-level complexity comparison for GLC-image on Kodak with resolution of 512 $\times$ 768. }
\centering
\resizebox{0.75\linewidth}{!}{
\begin{tabular}{c | c | c }
    \midrule
      Module & \makecell[c]{Running \\ time (ms)} & Params \\
    \midrule
       Latent encoder  & 27.3 & 29.4M \\
        \midrule
       Latent decoder  & 45.9 & 42.5M \\
    \midrule
          Transform encoder & 0.48 & 2.4M \\
        \midrule
        Transform decoder & 0.48 & 2.4M \\
        \midrule
\end{tabular}
}
  \label{tab:complexity_module}
  \vspace{-5mm}
\end{table}

\textbf{Complexity.} We conduct the complexity analysis for GLC-image and previous methods using an NVIDIA A100 Tensor Core GPU. As presented in Table \ref{tab:complexity_natural}, GLC-image not only show a better Rate-distortion performance measured by BD-Rate with DISTS (68.5\% performance improvement), but also can achieve and comparable latency compared to MS-ILLM on Kodak. Table~\ref{tab:complexity_module} presents the runtime and parameter count of GLC-image at the module level, including the latent encoder/decoder and the transform encoder/decoder. The results indicate that the majority of the computational cost is attributed to the latent encoder and decoder, while the transform encoder and decoder are lightweight and efficient.

\begin{figure*}[t]
  \centering
    \includegraphics[width=0.9\linewidth]{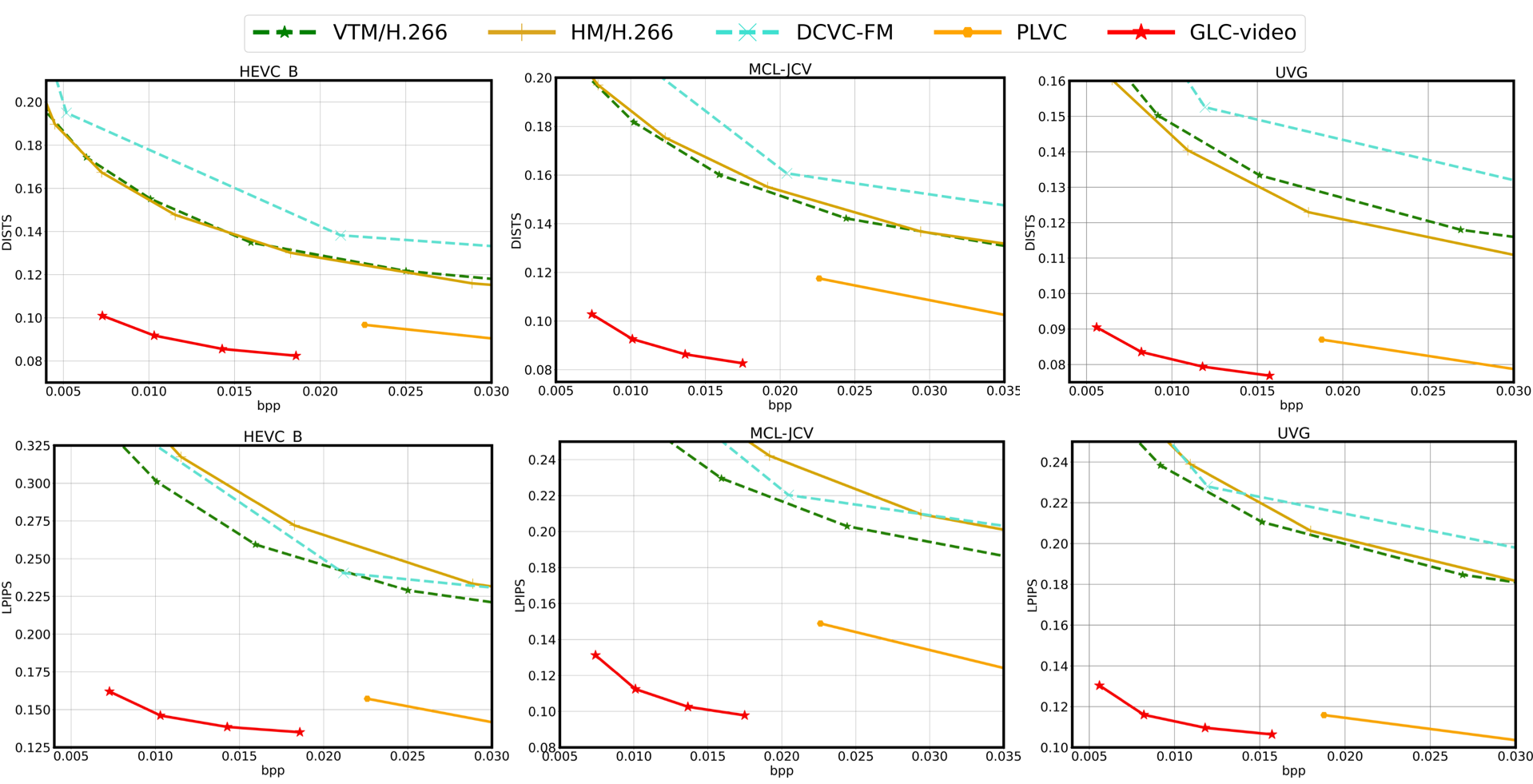}
    \vspace{-3mm}
    \caption{Rate-Distortion curves for comparing DISTS and LPIPS of GLC-video and other methods on HEVC class B, MCL-JCV, and UVG datasets.}
  \label{fig:glc-video-rd}
    \centering
    \vspace{2mm}
    \includegraphics[width=0.9\linewidth]{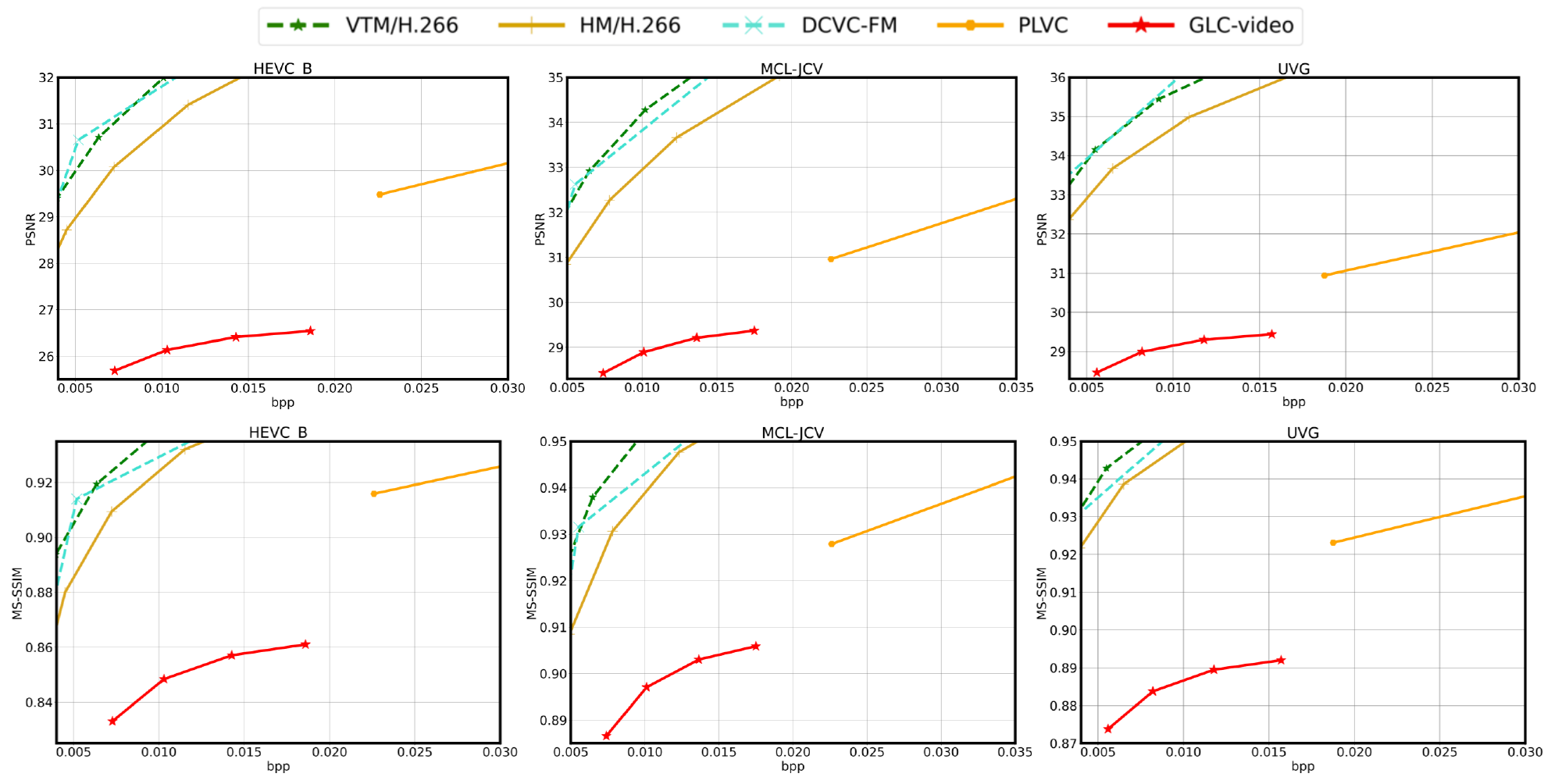}
    \vspace{-3mm}
    \caption{Rate-Distortion curves for comparing PSNR and MS-SSIM of GLC-video and other methods on HEVC class B, MCL-JCV, and UVG datasets.}
  \label{fig:glc-video-rd-psnr}
\end{figure*}

\begin{figure*}[t]
  \centering
  \includegraphics[width=\linewidth]{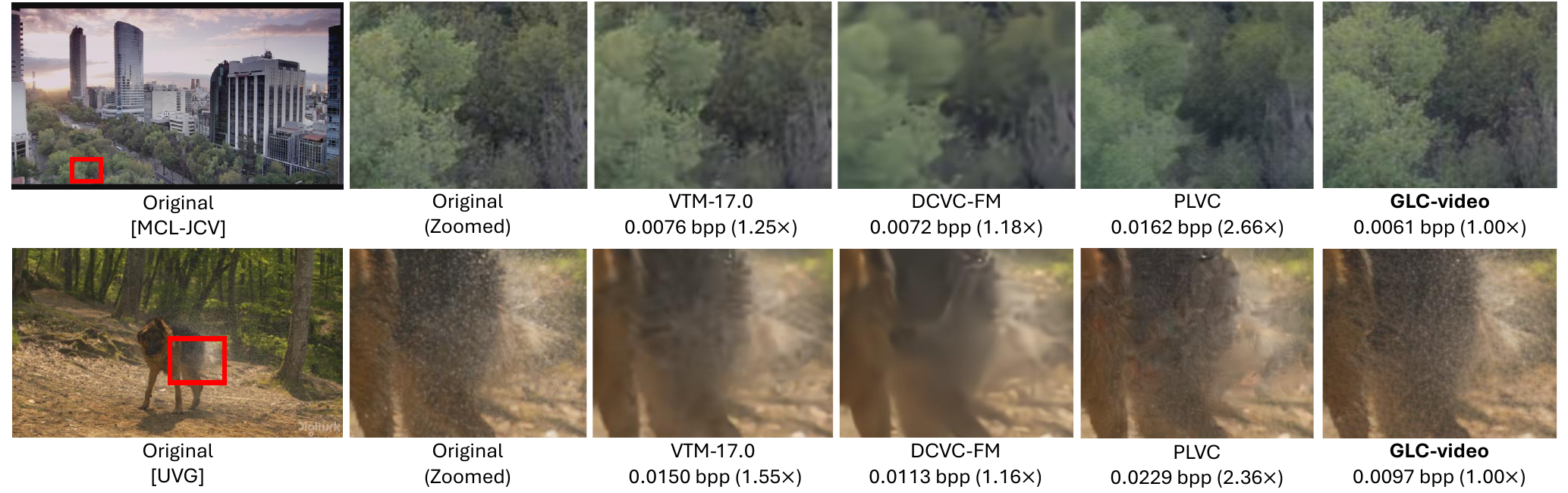}
  \vspace{-3mm}
    \caption{Qualitative examples for comparing the proposed GLC-video with other methods.  For each method, we not onwly annotate the bit per pixel (bpp) for coding the video sequence, but also show the bpp multiplier relative to our method.}
    \vspace{-3mm}
  \label{fig:VideoCompare}
\end{figure*}

\subsection{Experiments for Video Compression}
In this section, we present the experimental results of our GLC-video framework on benchmark datasets for video compression. Additionally, we conduct ablation studies to validate the effectiveness of the proposed spatio-temporal categorical hyper module. The performance is evaluated using BD-Rate~\cite{bjontegaard2001calculation} with respect to DISTS.

\begin{figure}[t]
  \centering
    \includegraphics[width=0.6\linewidth]{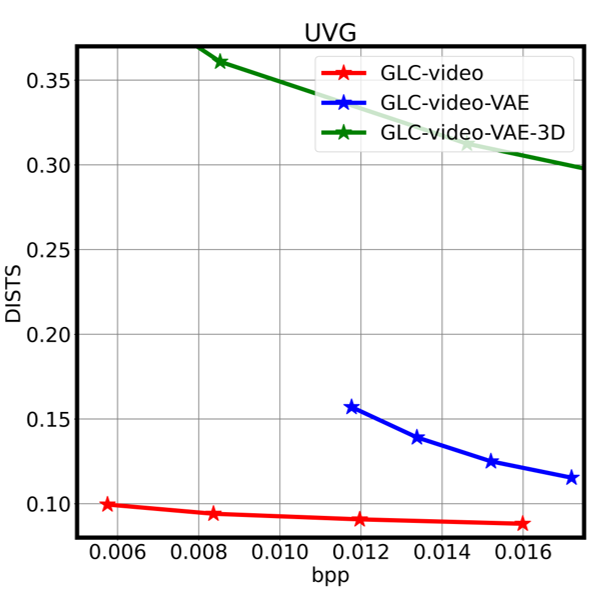}
      \vspace{-3mm}
    \caption{Ablation study on the latent encoder and decoder of GLC-video on UVG dataset (resized to 720p).}
  \label{fig:glc-video-latent-encdec}
    \vspace{-3mm}
\end{figure}

\textbf{Performance Comparisons.} In Fig.~\ref{fig:glc-video-rd}, the Rate-Distortion curves are illustrated for comparing our proposed GLC-video with HM/H.266, VTM/H.266~\cite{bross2021overview}, DCVC-FM~\cite{DCVC-FM} and PLVC~\cite{yang2022perceptual}. As mentioned, DISTS provides a more accurate assessment of visual quality and we mainly focus on this metric. Thanks to performing coding in the generative latent space, our proposed GLC-video can works at much lower bitrate than PLVC~\cite{yang2022perceptual}. Compared with traditional video codec HM/H.266, H.266/VTM, and the previous advanced neural codec DCVC-FM~\cite{DCVC-FM}, our method provides much better perceptual quality measured by DISTS and LPIPS. When using PLVC as the anchor method, GLC-video provides an average of 65.3\%  bitrate saving in term of DISTS on these video datasets, demonstrating the effectiveness of our method.

Fig.~\ref{fig:glc-video-rd-psnr} shows the rate-distortion curves in terms of PSNR and MS-SSIM for GLC-video. Similar to other generative codecs, GLC-video exhibits lower performance than non-generative methods under these distortion-based metrics. However, it is argued that perceptual metrics are more appropriate for evaluating the visual quality of our generative codec, especially at ultra-low bitrates. Although GLC-video produces reconstructions with lower PSNR and MS-SSIM scores, it can effectively reduce blurry artifacts and generate more visually appealing details (as shown in Fig.~\ref{fig:VideoCompare}).

\textbf{Visual Comparisons.} Fig.~\ref{fig:VideoCompare} presents a qualitative comparison of our GLC-video against other methods. At ultra-low bitrates, H.266/VTM, HM/H.266, and DCVC-FM exhibit limited perceptual quality, characterized by noticeable blurry artifacts. While PLVC achieves enhanced perceptual quality through the use of GAN loss, our GLC-video leverages the generative latent space of VQ-VAE, which aligns more closely with human perception. As a result, GLC-video produces more visually appealing reconstructions, even at bitrates lower than those used by PLVC.

\begin{figure}[t]
  \centering
    \includegraphics[width=\linewidth]{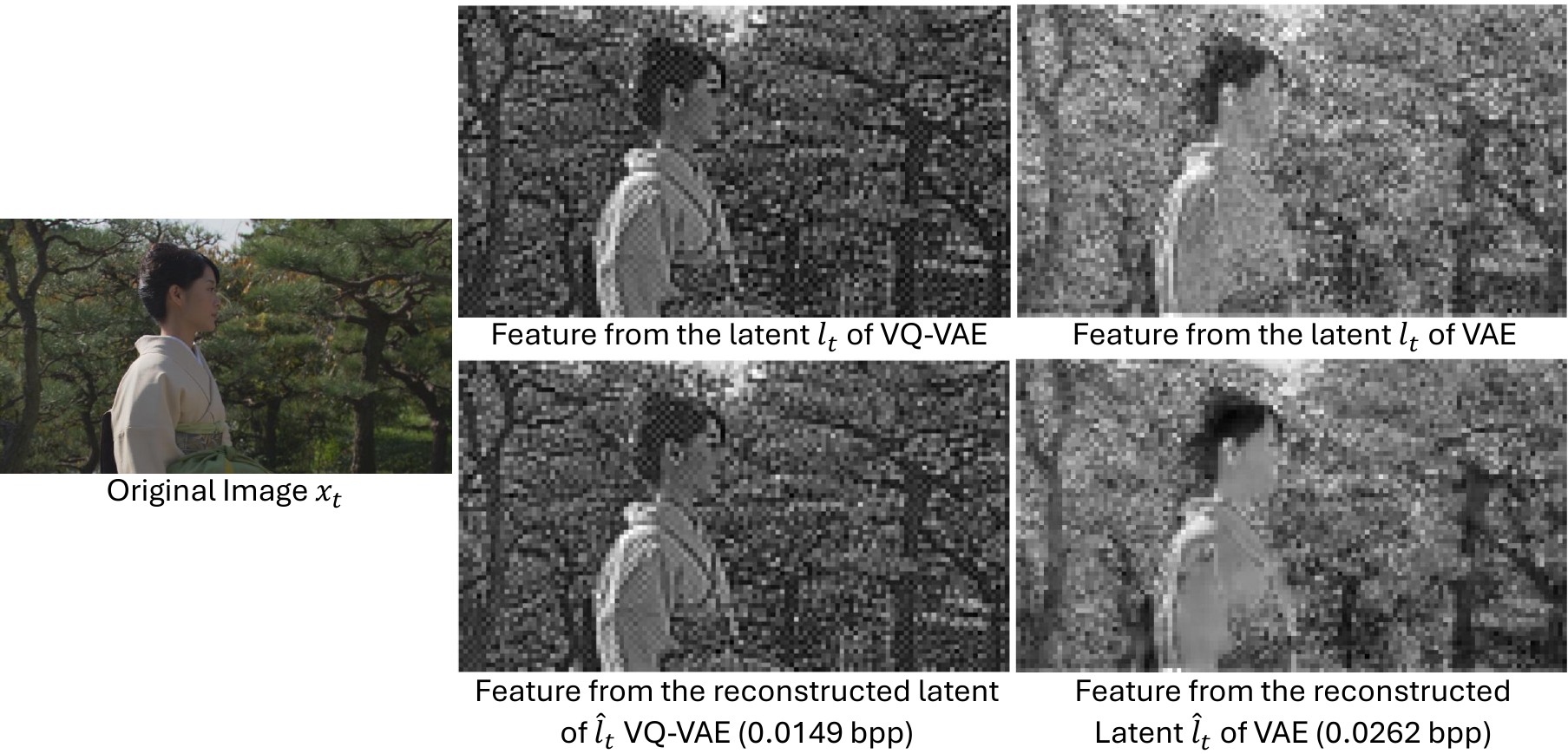}
      \vspace{-3mm}
    \caption{Visualization for comparing the latent of VQ-VAE and VAE.
    The feature map of VQ-VAE is more distinct than that of the VAE. Its reconstruction is also sharper, while VAE shows noticeable blurring (even at higher bitrates).
    }
  \label{fig:abl-feature-visual}
    \vspace{-3mm}
\end{figure}

\textbf{Ablation on the Latent Auto-Encoder.}
Temporal redundancy can also be reduced by using a 3D VAE to jointly compressing spatial and temporal information. However, a smaller latent dimension does not guarantee a lower bitrate, as the latent may still retain high entropy.
Figure~\ref{fig:glc-video-latent-encdec} shows a comparison of latent auto-encoders for GLC-video on the UVG dataset (downsampled to 720p via bicubic interpolation to save memory). We adopt the 3D VAE from CogVideoX~\cite{yang2024cogvideox} as the latent auto-encoder, denoted as GLC-video-VAE-3D. For fairness, we also implement GLC-video-VAE, which uses a 2D VAE with a similar parameter size to the VQ-VAE used in GLC-video.
GLC-video-VAE-3D performs worse than GLC-video-VAE, mainly due to higher entropy in its latent representation despite lower dimensionality. Although off-the-shelf 3D video tokenizers are still limited, joint spatial-temporal modeling remains a promising direction for further exploration.
Moreover, replacing VQ-VAE with a standard VAE results in noticeable performance degradation, primarily because the VQ-VAE produces a sparser latent space. As shown in Fig.~\ref{fig:abl-feature-visual}, VQ-VAE retains semantic information more effectively under ultra-low bitrate settings.

\begin{table}[t]
    \caption{Ablation study on latent-space compression for GLC-video.}
    \centering
    \resizebox{\linewidth}{!}{
    \begin{tabular}{c | c | c}
        \midrule
        Latent coding scheme & Hyper module & BD-Rate $\downarrow$ \\
        \midrule
        \makecell[c]{w/o conditional \\ coding}   & \makecell[c]{Spatial \\ categorical based} & 214.6\%\\
        \midrule
        \multirow{3}{*}{}      
            & \makecell[c]{Factorized \\ based}  & 20.7\%\\
        \cmidrule(lr){2-3}
          \textbf{\makecell[c]{w/ conditional \\ coding}}  & \makecell[c]{Spatial \\ categorical based} & 0\% \\   
        \cmidrule(lr){2-3} 
            & \textbf{\makecell[c]{Spatio-temporal \\ categorical based}} & \textbf{--22.5\%} \\
        \midrule
    \end{tabular}
    }
    \label{tab:ablation-video-hyper}
      \vspace{-3mm}
\end{table}

\begin{table}[t]
    \caption{Ablation study on the training strategy for GLC-video.}
    \centering
    \resizebox{0.95\linewidth}{!}{
    \begin{tabular}{c | c | c}
        \midrule
        Training strategy & \makecell[c]{w/o joint training} &  \makecell[c]{\textbf{w/ joint training}}\\
      \midrule      
      BD-Rate $\downarrow$  & 0 &\textbf{--49.8\%} \\
        \midrule
    \end{tabular}
    }
    \label{tab:ablation-video-joint-train}
    \vspace{3mm}
      \caption{Ablation Study on the number of tokens $K$ of the spatio-temporal categorical hyper module of GLC-video.}
    \centering
    \resizebox{0.95\linewidth}{!}{
        \tabcolsep=4pt
	\begin{tabular}{c | c c c c c}
	    \midrule
		 \# Tokens & $K=4$ & $K=8$  & $\textbf{K=16}$ & $K=24$ & $K=32$ \\
	    \midrule
            BD-Rate $\downarrow$ & 13.0\% & --3.9\%  & \textbf{--22.5\%} &  --10.1\% & --10.0\%\\
	    \midrule
	\end{tabular}
	}
  \label{tab:ablation-video-tokens}
\end{table}

\textbf{Ablation Study on conditinal coding.} 
As shown in Table~\ref{tab:ablation-video-hyper}, introducing conditional coding significantly improves overall performance. This improvement stems from the fact that compressing the current latent benefits from temporal context extracted from the previously decoded latents, effectively reducing temporal redundancy. During stage II (transform coding learning) and stage III (joint training), feature propagation further enhances the utilization of correlations across multiple frames.

\textbf{Ablation Study on the Hyper Module.} Table~\ref{tab:ablation-video-hyper} also compares the proposed spatio-temporal categorical hyper module with the factorized based hyper module and the spatial categorical hyper module from GLC-image. The spatial categorical hyper module demonstrates better performance than the factorized based approach, consistent with observations in GLC-image for image compression. Notably, the proposed spatio-temporal categorical hyper module achieves the best performance, providing a 22.5\% bitrate saving compared to the spatial categorical hyper module. This improvement can be attributed to the incorporation of temporal context and the effective exploitation of the sparsity along the temporal dimension in hyper-information. By representing $z_t$ more compactly, our method significantly reduces the bitrate.

\textbf{Ablation Study on the Number of Tokens $K$.} $K$ controls how many tokens are computed as $z_t$ for transmitting the hyper information, thus influencing the capacity of the proposed spatio-temporal categorical hyper module and the corresponding bit cost.
Table~\ref{tab:ablation-video-tokens} presents a comparison of performance for different values of $K$.
We use the model equipped with the spatial categorical hyper module as the anchor method. A smaller $K$(e.g., $K=4$ and $K=8$) may fail to effectively represent the hyper-information, leading to a performance drop. Conversely, a larger $K$ increases the bit cost for representing the hyper-information and adds complexity to learning the attention maps $M_t$, thereby increasing training difficulty. The results show that $K=16$ achieves the best performance, and this value is adopted in our model.

\begin{figure}[t]
  \centering
    \includegraphics[width=\linewidth]{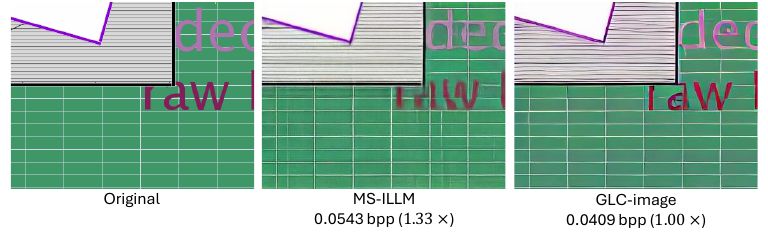}
      \vspace{-5mm}
    \caption{Visual comparison for GLC-image on a screen image. Discrepancies from the original image are observed in the text region.}
  \label{fig:limitations}
  \vspace{-3mm}
\end{figure}

\section{Limitations}
Despite the improvements introduced, GLC-image and GLC-video still encounter several limitations. Although the proposed methods maintain semantic-level consistency and enhance visual quality, they still show limitations in pixel-level metrics such as PSNR, similar to existing generative approaches. GLC-image struggles with accurately reconstructing texts, as evident in the distortions shown in Fig.~\ref{fig:limitations}. For GLC-video, several techniques have been adopted to enhance temporal consistency: 1. feature propagation during training and 2. a two-frame input to the discriminator for Patch-GAN adversarial loss. However, though GLC-video show more stable reconstructions than PLVC~\cite{yang2022perceptual}, it still falls short of the non-generative method DCVC-FM~\cite{DCVC-FM} in terms of pixel-level stability. Flickering artifacts can still be observed in some regions under close inspection.

\section{Conclusion}

In this paper, we introduce a generative latent coding (GLC) scheme for image compression and video compression, referred to as GLC-image and GLC-video, respectively.
Unlike most existing pixel-space codecs, GLC-image and GLC-video perform transform coding in the latent space of a generative VQ-VAE, achieving high-fidelity and high-realism generative compression at ultra-low bitrates. 
Furthermore, compared to VQ-indices-map coding, our approach supports rate-variable compression, a critical feature for practical image and video codecs. GLC-image investigates to code images in the generative latent space and incorporate a spatial categorical hyper module to reduce the bit cost of coding hyper information. GLC-video further verifies the feasibility of coding videos in the generative latent space and proposes a spatio-temporal categorical hyper module to non-uniformly encode the hyper information to improve the performance. To fully leverage the potential of such generative latent coding pipeline, A code-prediction-based supervision is introduced to facilitate the training.
Experimental results show that GLC-image and GLC-video achieve higher perceptual quality at much lower bitrates compared to existing generative codecs.

\bibliographystyle{IEEEtran}
\bibliography{glc-bib}




\vspace{-35pt}
\begin{IEEEbiography}[{\includegraphics[width=1in,height=1.25in,clip,keepaspectratio]{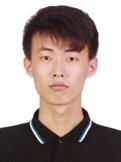}}]{Linfeng Qi} received the B.S. degree in electronic information engineering from University of Science and Technology of China (USTC), Hefei, Anhui, China, in 2020. He is currently pursuing the Ph.D. degree with the Department of Electronic Engineering and Information Science, USTC. He is also an intern at the Media Computing Group at Microsoft Research Asia. His research interests include video compression and media computing.
\end{IEEEbiography}

\vspace{-40pt}
\begin{IEEEbiography}[{\includegraphics[width=1in,height=1.25in,clip,keepaspectratio]{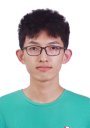}}]{Zhaoyang Jia} received his B.S. degree in 2022 from the University of Science and Technology of China (USTC). Currently, he is pursuing the Ph.D. degree at the University of Science and Technology of China. He is also an intern at the Media Computing Group at Microsoft Research Asia. His research interests include media compression, media computing and digital watermarking.
\end{IEEEbiography}
 
\vspace{-40pt}
\begin{IEEEbiography}[{\includegraphics[width=1in,height=1.25in,clip,keepaspectratio]{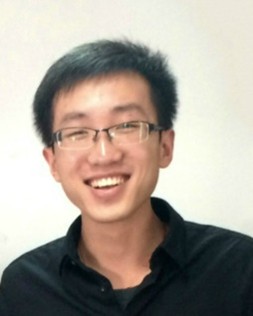}}]{Jiahao Li} received the B.S. degree in computer science and technology from the Harbin Institute of Technology in 2014, and the Ph.D. degree from Peking University in 2019. He is currently a Senior Researcher with the Media Computing Group, Microsoft Research Asia. His research interests include neural video compression and other video tasks, like video backbone design and video representation learning. He has more than thirty published papers, standard proposals, and patents in the related area.
\end{IEEEbiography}

\vspace{-40pt}
\begin{IEEEbiography}[{\includegraphics[width=1in,height=1.25in,clip,keepaspectratio]{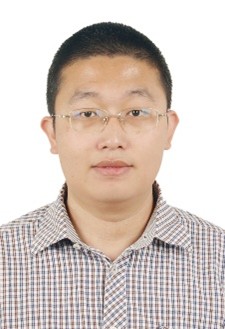}}]{Bin Li} (Member, IEEE) received the B.S. and Ph.D. degrees in electronic engineering from the University of Science and Technology of China (USTC), Hefei, Anhui, China, in 2008 and 2013, respectively. He joined Microsoft Research Asia (MSRA), Beijing, China, in 2013 and now he is a Principal Researcher. He has authored or co-authored over 50 papers. He holds over 30 granted or pending U.S. patents in the area of image and video coding. He has more than 40 technical proposals that have been adopted by Joint Collaborative Team on Video Coding. His current research interests include video coding, processing, transmission, and communication. Dr. Li received the best paper award for the International Conference on Mobile and Ubiquitous Multimedia from Association for Computing Machinery in 2011. He received the Top 10\% Paper Award of 2014 IEEE International Conference on Image Processing. He received the best paper award of IEEE Visual Communications and Image Processing 2017.
\end{IEEEbiography}

\vspace{-40pt}
\begin{IEEEbiography}[{\includegraphics[width=1in,height=1.25in,clip,keepaspectratio]{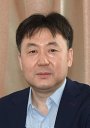}}]{Houqiang Li} (Fellow, IEEE) received the B.S., M.Eng., and Ph.D. degrees in electronic engineering
from the University of Science and Technology of China, Hefei, China, in 1992, 1997, and 2000, respectively. He is a Professor with the Department of Electronic Engineering and Information Science, University of Science and Technology of China. He was a Winner of the National Science Funds (NSFC) for Distinguished Young Scientists, the Distinguished Professor of the Changjiang Scholars Program of China, and the Leading Scientist of the Ten Thousand Talent Program of China. He has authored or coauthored over 200 papers in journals and conferences. His research interests include multimedia search, image/video analysis, video coding, and communication. He was a recipient of the National Technological Invention Award of China (second class) in 2019 and the National Natural Science Award of China
(second class) in 2015. He was also a recipient of the Best Paper Award for VCIP 2012, the Best Paper Award for ICIMCS 2012, and the Best Paper Award for ACM MUM in 2011. He served as the TPC Co-Chair for VCIP 2010 and the General Co-Chair for ICME 2021. He served as an Associate Editor for IEEE TRANSACTIONS ON CIRCUITS AND SYSTEMS FOR VIDEO TECHNOLOGY, from 2010 to 2013.
\end{IEEEbiography}

\vspace{-40pt}
\begin{IEEEbiography}[{\includegraphics[width=1in,height=1.25in,clip,keepaspectratio]{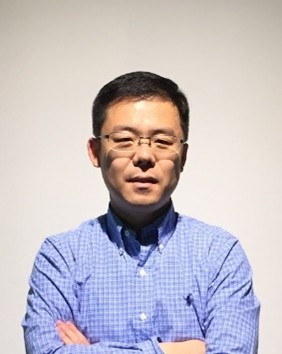}}]{Yan Lu} received his Ph.D. degree in computer science from Harbin Institute of Technology, China. He joined Microsoft Research Asia in 2004, where he is now a Partner Research Manager and manages research on media computing and communication. He and his team have transferred many key technologies and research prototypes to Microsoft products. From 2001 to 2004, he was a team lead of video coding group in the JDL Lab, Institute of Computing Technology, China. From 1999 to 2000, he was with the City University of Hong Kong as a research assistant. Yan Lu has broad research interests in the fields of real-time communication, computer vision, video analytics, audio enhancement, virtualization, and mobile-cloud computing. He holds 30+ granted US patents and has published 100+ papers in refereed journals and conference proceedings.

\end{IEEEbiography}

\end{document}